\documentclass[twocolumn,prb,nofootinbib,superscriptaddress,12]{revtex4-1}
\usepackage{color,times,url,graphicx,afterpage}
\usepackage{multirow,bm,braket,soul}
\usepackage{amsmath,xcolor,hyperref}
\usepackage[normalem]{ulem}	
\usepackage{hyperref,footnote}

\begin{document}

\title{The geometric blueprint of perovskites}

\author{Marina R. Filip}
\affiliation{Department of Materials, University of Oxford, Parks Road, Oxford OX1 3PH, United 
Kingdom}
\author{Feliciano Giustino}
\email{feliciano.giustino@materials.ox.ac.uk}
\affiliation{Department of Materials, University of Oxford, Parks Road, Oxford OX1 3PH, United 
Kingdom}
\affiliation{Department of Materials Science and Engineering, Cornell University, Ithaca, New York 
14853, USA}

\keywords{$|$ Perovskites$ | $Structure prediction $|$ Goldschmidt $|$ Data mining $|$} 

\begin{abstract}
 Perovskite minerals form an essential component of the Earth's mantle, and synthetic crystals
are ubiquitous in electronics, photonics, and energy technology. The extraordinary chemical
diversity of these crystals raises the question on how many and which perovskites are yet to be
discovered. Here we show that the `no-rattling' principle postulated by Goldschmidt in 1926,
describing the geometric conditions under which a perovskite can form, is much more effective than
previously thought, and allows us to predict new perovskites with a fidelity of 80\%. By
supplementing this principle with inferential statistics and internet data mining we establish that
currently known perovskites are only the tip of the iceberg, and we enumerate ninety thousand
hitherto-unknown compounds awaiting to be studied.
Our results suggest that geometric blueprints may enable the systematic screening of millions of
compounds, and offer untapped opportunities in structure prediction and materials design.
\end{abstract}

\maketitle

\section{Introduction}

Crystals of the perovskite family rank among the most common ternary and quaternary 
compounds and are central to many areas of current research\cite{Muller-Roy}. For example silicate 
perovskites constitute the most abundant minerals on Earth\cite{Murakami2012}, and synthetic oxide 
perovskites find applications as ferroelectrics\cite{Benedek2013}, ferromagnets\cite{Loudon2002},
multiferroics\cite{Mundy2016}, high-temperature superconductors\cite{Bednorz1986}, magnetoresistive
sensors\cite{Milward2005}, spin filters\cite{Park1997}, superionic conductors\cite{Navrotsky1999}
and catalysts\cite{Suntivich2011}. Halide perovskites are promising for high-efficiency solar cells,
light-emitting diodes and lasers\cite{Zhang2016,Tan2014,Zhu2015}; their double perovskite
counterparts are efficient scintillators for radiation detection\cite{Giustino2016}. The unique
versatility of the perovskite crystal structure stems from its unusual ability to accommodate a
staggering variety of elemental combinations. This unparalleled diversity raises the questions on
how many new perovskites are yet to be discovered, and which ones will exhibit improved or novel
functionalities. In an attempt to answer these questions, we here begin by mapping the entire
compositional landscape of these crystals.

Figure~\ref{fig1}a shows the structure of a cubic ABX$_3$ perovskite. In this structure the A and B
elements are cations and X is an anion. B-site cations are six-fold coordinated by anions to form
BX$_6$ octahedra. The octahedra are arranged in a three-dimensional corner-sharing network, and each
cavity of this network is occupied by one A-site cation\cite{Megaw}. All perovskites share the same
network topology, but can differ in the degree of tilting and distortions of the 
octahedra\cite{Megaw, Glazer1972, Woodward1997-1, Woodward1997-2, Lufaso2001}. The quaternary 
counterpart of the perovskite crystal is the double perovskite A$_2$BB$'$X$_6$. When the B and 
B$^\prime$ cations alternate in a rock-salt sublattice, the crystal is called 
elpasolite\cite{Giustino2016} (Figure~S1a). In the following we use the term `perovskite' to 
indicate both ternary and quaternary compounds.

\section*{Results}
How many perovskites do currently exist? If we search for the keyword `perovskite' in the inorganic
crystal structure database (ICSD) we find 8866 entries\cite{icsd}, but after removing duplicates the
headcount decreases to 335 distinct ABX$_3$ compounds. Similarly, the keyword `elpasolite' yields
224 distinct A$_2$BB$'$X$_6$ compounds. By including also the extensive compilations of Refs.
\citenum{Flerov1998,Meyer1982,Li2004,Li2008,Vasala2015,Galasso,Flerov2001}, we obtain a grand total
of 1622 distinct crystals that are reliably identified as perovskites (database~S1.1).

How many perovskites are left to discover? Direct inspection of the elemental composition in
database~S1.1 indicates that, with the exception of hydrogen, boron, carbon, phosphorous and some
radioactive elements, these crystals can host every atom in the Periodic Table. Therefore an upper
bound for the number of possible perovskites is given by all the combinations of three cations and
one anion. By considering only ions with known ionic radii\cite{Shannon} we count 3,658,527
hypothetical compounds. Our goal is to establish which of these compounds can form perovskite
crystals. Ideally we should like to resort to {\it ab initio} computational
screening\cite{Curtarolo2013}, but these techniques are not yet scalable to millions of compounds.
For example, by making the optimistic assumption that calculating a phase diagram required only one
hour of supercomputing time per compound, it would take 160~years to complete this task.

An empirical approach to investigate the formability of ABX$_3$ perovskites was proposed by
Goldschmidt almost a century ago\cite{Goldschmidt}. In this approach the perovskite structure is
described as a collection of rigid spheres, with sizes given by the ionic radii $r_{\rm A}$,
$r_{\rm B}$ and $r_{\rm X}$. These radii can be combined into two dimensionless descriptors, the
tolerance factor $t=(r_{\rm A}+r_{\rm X})/\sqrt{2}(r_{\rm B}+r_{\rm X})$, and the octahedral factor
$\mu = r_{\rm B}/r_{\rm X}$. Goldschmidt postulated that perovskites arrange so that {\it `the
number of anions surrounding a cation tends to be as large as possible, subject to the condition
that all anions touch the cation'} \cite{Megaw}. This statement constitutes the `no-rattling'
principle, and limits the range of values that $t$ and $\mu$ can take for a perovskite.

Goldschmidt's principle was recently tested on larger datasets than those available in 1926. By 
analysing a few hundred ternary oxides and halides, it was found that in a two-dimensional $t$ vs.
$\mu$ map perovskites and non-perovskites tend to cluster in distinct regions\cite{Li2004, Li2008,
Pilania2016}. Based on this observation, much work has been dedicated to identifying a 
`stability range', 
either by postulating boundaries for $t$ and $\mu$\cite{Li2004,Li2008,Giaquinta1994}, or by using 
machine learning to draw $t$~vs.~$\mu$ curves enclosing the data points\cite{Pilania2016}. The merit 
of these efforts is that they addressed the predictive power of the no-rattling principle in 
qualitative terms.
However, these approaches suffer from relying too heavily on empirical ionic 
radii. It is well known that the definition of ionic radii is non-unique, and that even within the 
same definition there are variations reflecting the coordination and local 
chemistry\cite{Muller-Roy,Megaw,Goldschmidt,Pauling, Shannon}. As these uncertainties transfer to 
the octahedral and tolerance factors, the stability ranges proposed so far are descriptive rather 
than predictive. In order to overcome these limitations, instead of defining a map starting from 
empirical data, our strategy is to construct a stability range from first principles, by relying 
uniquely on Goldschmidt's hypothesis. This choice allows us to also derive a structure map for 
quaternary compounds, a step that has thus far remained elusive.

\begin{figure*}[t!]
\linespread{1.2}
\centering
\includegraphics[width=0.85\textwidth]{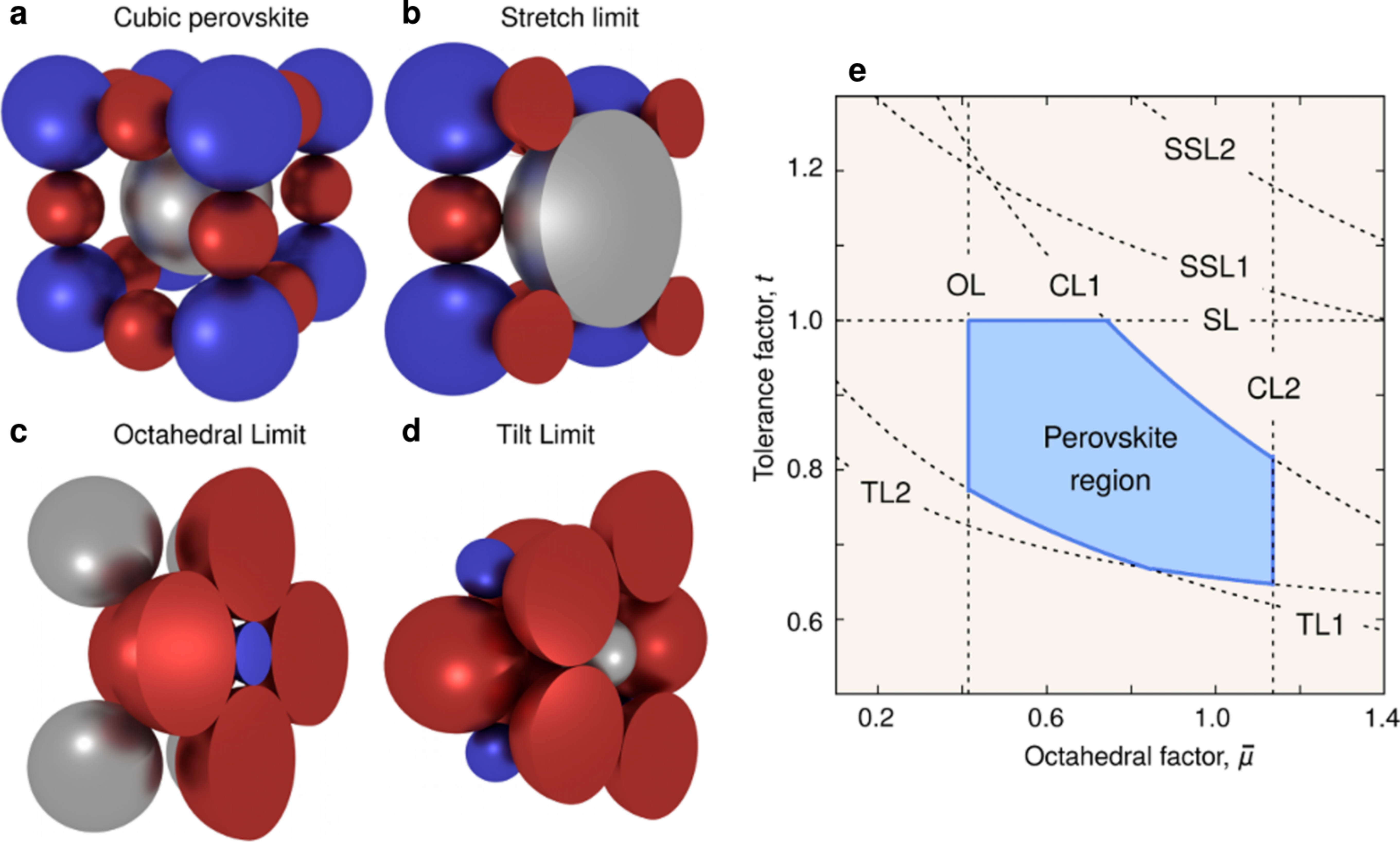}
\caption{\small  
  {\bf No-rattling principle for ternary perovskites}.
  {\bf a}, Rigid-sphere representation of the conventional cell of a cubic ABX$_3$
  perovskite, with A in grey, B in blue, and X in red. {\bf b}, Cross-sectional view of
  the stretch limit. In this configuration the A cation sits at the center of the cavity
  and touches twelve nearest-neighbor X anions. {\bf c}, Cross-section of the perovskite
  structure in the octahedral limit. Here the nearest-neighbor X anions belonging to the
  same octahedron touch. {\bf d}, Schematic representation of the tilt limit. This
  configuration can be thought of as obtained from {\bf b} by reducing the size of the A
  cation (grey), increasing the size of the X anion (red), and tilting the octahedra so
  that anions of adjacent octahedra touch. In this case the A cation moves away from the
  center of the cavity so as to optimally fill the available space, and touches five X
  anions. {\bf e}, Stability area of ternary perovskites (blue), as derived from the
  no-rattling principle and from additional chemical limits (dashed lines): stretch limit
  (SL), octahedral limit (OL), tilt limits (TL1 and TL2), chemical limits (CL1 and CL2),
  secondary stretch limits (SSL1 and SSL2). These boundaries are derived in the
  Supplementary Information.
\label{fig1}}
\end{figure*}

We describe our strategy starting from ternary perovskites, and then we generalize our findings to
quaternary perovskites.  Figure~\ref{fig1}b shows that, for the A cation to fit in the cavity, the
radii must satisfy the condition $r_{\rm A} + r_{\rm X} \le \sqrt{2}(r_{\rm B} + r_{\rm X})$, or
equivalently $t\leq1$ (see Supplementary Information). Similarly, Figure~\ref{fig1}c shows that the
octahedral coordination of the B cation by six X anions is not possible when $\sqrt{2}(r_{\rm B} + 
r_{\rm X} ) < 2r_{\rm X}$, therefore $\mu \geq \sqrt{2} - 1$. We refer to these conditions as the
`octahedral' limit and the `stretch' limit, respectively, as shown in Figure~\ref{fig1}e. When these
conditions are not fulfilled, the lattice tends to distort towards a layered geometry, with
edge-sharing or face-sharing octahedra, or lower B-site coordination\cite{Megaw, Giaquinta1994}. 
These two bounds are well-known\cite{Megaw, Goldschmidt} but are insufficient for quantitative 
structure prediction.

When t is smaller than 1, the corner-sharing octahedra exhibit an increased degree of 
tilting, and the A-site cation is displaced from the central position in the cuboctahedral cavity, 
as shown Figures~\ref{fig1}d and S2\cite{Woodward1997-1, Woodward1997-2}. Previous work has shown 
that the displacement of the A-site cation can be determined by optimizing Coulomb interactions 
between the perovskite ions, taking into account the bond-valence configuration of each 
ion\cite{Lufaso2001}. Here we explore this scenario using a purely geometric approach, by identifying 
the ionic positions which achieve the tightest packing of ions in a tilted perovskite configuration.

Figure~\ref{fig1}d shows that the 
octahedra can tilt only until two anions belonging to adjacent octahedra come into contact. In this 
extremal configuration, to satisfy the no-rattling principle the A-site cation must exceed a 
critical size. Using the  geometric construction described in the Supplementary Information 
(Figure~S2), this condition translates into a lower limit for the tolerance  factor,
$t\geq \rho_\mu/\sqrt{2} (\mu+1)$,
where $\rho_\mu$ is a simple piece-wise linear function of $\mu$ (Figure~S3) (see Supplementary
Information). We refer to this condition as the `tilt' limit (Figure~\ref{fig1}e). When this
criterion is not met, the perovskite network tends to collapse into structures with edge-sharing or
face-sharing octahedra. Along the same lines we must consider the limit of two neighboring A-site
cations coming into contact (Figure~\ref{fig1}e and Figure~S4a), and of the A and B cations touching
(Figure~S4b). Besides these geometric constraints, we also ought to take into account that the rigid
spheres of the model represent chemical elements, therefore we have additional bounds on the size of
the ionic radii: the largest tolerance factor corresponds to the combination of Cs and F, and the
oxidation number of A cannot exceed that of B according to Pauling's valency rule\cite{Pauling}. By
considering all the $(\mu,t)$ points that satisfy these conditions simultaneously, we obtain the
perovskite stability area shown in blue in Figure~\ref{fig1}e.

\begin{figure*}[t!]
\linespread{1.2}
\centering
\includegraphics[width=0.85\textwidth]{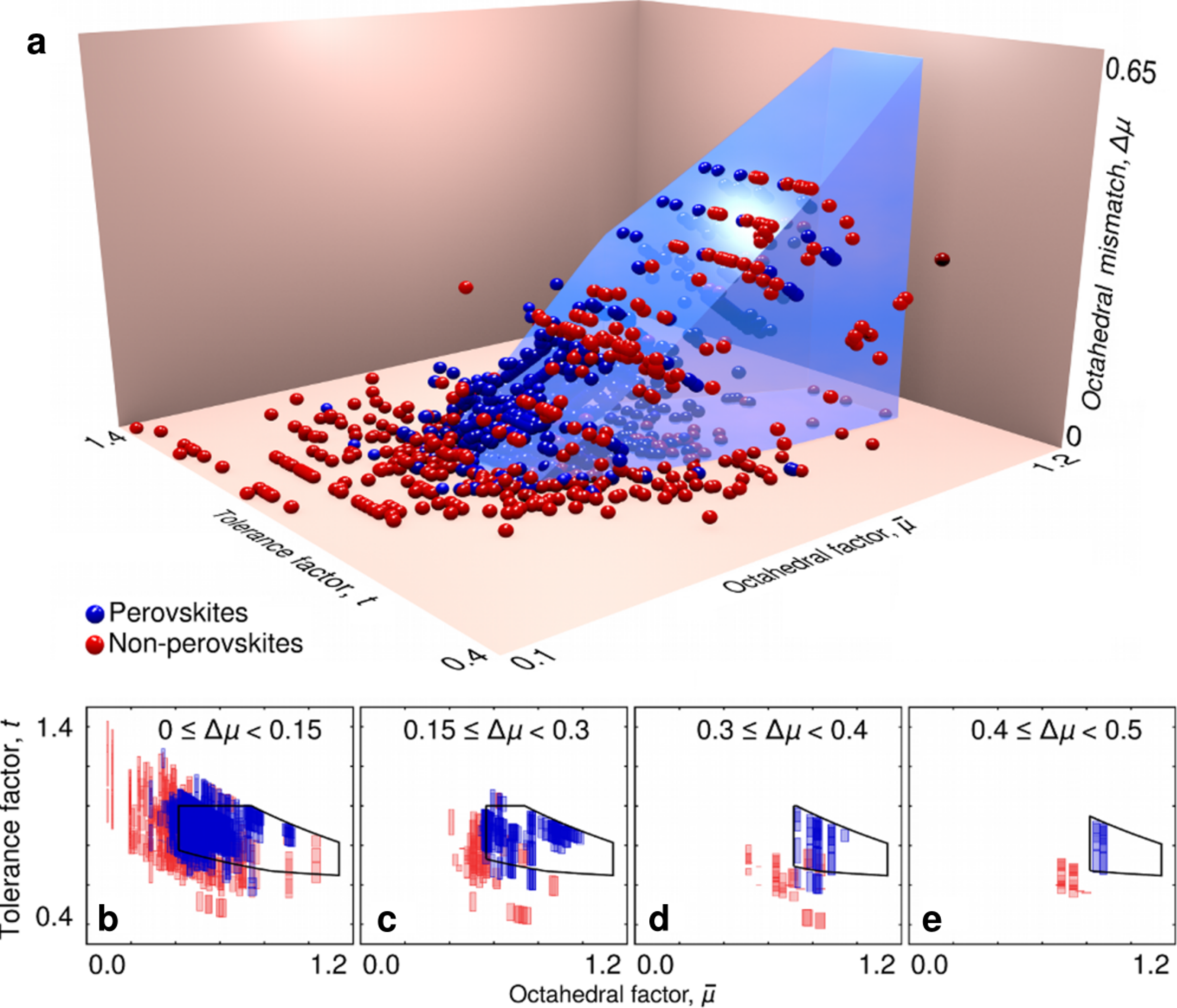}
\caption{{\bf Stability range of ternary and quaternary perovskites}.
  {\bf a}, The blue volume represents the stability range of perovskites in the
  $(\bar{\mu},t,\Delta\mu)$ space, as derived from Goldschmidt's no-rattling principle.
  The blue and red markers correspond to perovskites (databases S1.1 and S1.3) and
  non-perovskites (database S1.2), respectively, calculated for all the compounds in
  database~S1. The location of each marker is the center of the rectangular cuboid defined
  in the Supplementary Information.
  {\bf b} to {\bf e}, Slices of the stability volume shown in {\bf a}, reporting all
  perovskites with octahedral mismatch $\Delta\mu$ in the range indicated at the top of
  each panel. In these two-dimensional representations the cuboids appear as rectangles,
  with blue and red indicating perovskites and non-perovskites, respectively.
\label{fig2} }
\end{figure*}

These considerations are readily generalized to double perovskites. In this case we have two 
different cations B and B$^\prime$ (Figure~S1a), therefore we must consider two octahedral
parameters: the average octahedral factor, $\bar{\mu} = (r_{{\rm B}}+r_{{\rm B}'})/2\,r_{\rm X}$,
and the octahedral mismatch $\Delta\mu = |r_{{\rm B}}-r_{{\rm B}'}|/2\,r_{\rm X}$. In the
Supplementary Information we derive the generalized tolerance factor, which takes the form $t = 
(r_{\rm A}/r_{\rm X}+1) /[2(\bar{\mu}+1)^2+ \Delta\mu^2]^{1/2}$. As we now have three structure
descriptors, the generalized stability region is a closed volume in the $(\bar{\mu},t,\Delta\mu)$
space, as seen in Figure~\ref{fig2}a. If we slice this volume through the plane $\Delta\mu = 0$ we
revert to the perovskite area of Figure~\ref{fig1}d. We emphasize that our present results derive
exclusively from Goldschmidt's principle (barring the chemical limits which are not essential), and
make no reference to the definition and values of the ionic radii. The bounds of the perovskite
regions are easy to evaluate for any structure, and are described by six inequalities for
$\bar{\mu}$, $t$ and $\Delta\mu$ in Table~S1.

Can the inequalities in Table~S1 be used for structure prediction? To answer this question we 
analyzed a record of 2291 ternary and quaternary compounds that we collected from the ICSD and from
Refs.~\citenum{Flerov1998,Meyer1982,Li2004,Li2008,Vasala2015,Galasso,Flerov2001}. This dataset
includes 1622 perovskites (database S1.1), 592 non-perovskites (database S1.2), and 77 compounds
which can crystallize either as a perovskite or another structure (database S1.3; see Supplementary
Information). Figure~\ref{fig2}a shows the distribution of all these compounds in the $(\bar{\mu},t,
\Delta\mu)$ space. We see that our perovskite volume (blue) delimits remarkably well the regions
occupied by perovskites (blue markers) and non-perovskite (red markers). A detailed view of these
data is provided in Figure~\ref{fig2}b--e, where we show slices of the perovskite volumes at fixed
octahedral mismatch, and in Figure~S5, where data for perovskites and non-perovskites are presented
separately. In these panels we see that, as $\Delta\mu$ increases, the stability region decreases in
size and moves towards higher octahedral factors. Remarkably, most datapoints from database~S1.1
closely follow this trend. Case-by-case inspection reveals several outliers, that is perovskite
markers falling outside of the perovskite region or vice-versa. The existence of outliers is to be
expected given the simplicity of Goldschmidt's model, but intriguingly we find many cases where
the presence of outliers signals the occurrence of polymorphism. An important example is BaTiO$_3$:
while this compound is mostly known as a ferroelectric perovskite, it is also stable in a hexagonal
structure under the same pressure and temperature conditions\cite{Akimoto1994}.

We now assess the predictive power of the model on quantitative grounds. The simplest way to proceed
would be to classify compounds based on whether the corresponding $(\bar{\mu},t,\Delta\mu)$ point
falls inside or outside the stability region. However, this procedure is 
unreliable as it is very sensitive to small variations in the ionic radii. A better strategy is to 
replace each point by a rectangular cuboid, with dimensions representing the uncertainty in the 
ionic radii. The uncertainty calculation is detailed in the Supplementary Information. With this 
choice we define the `formability' as the fraction of the cuboid volume falling within the perovskite 
region, and we classify the compound as a perovskite if this fraction exceeds a critical value 
(Figure~S6a). In order to quantify the accuracy of this classification procedure and the associated 
uncertainty, we determine the classification of large subsets of compounds, randomly selected from 
databases~S1.1 and S1.2, and we repeat this operation several thousand times.  By the central limit 
theorem, the average success rates tend to a normal distribution (Figure~S7b); the center of this 
distribution gives the most probable success rate, and the standard deviation yields the statistical 
uncertainty (Figure~S6c).

Our main result is that, for sample sizes of 100 compounds or more, the geometric model correctly
classifies 79.7$\pm$4.0\% of all compounds with a 95\% confidence level. This predictive power is
unprecedented among structure prediction algorithms.

For completeness we also assess how our model compares with previous models. To this 
end, we calculate how many of the known compounds in Databases S1.1 and S1.2 are correctly 
classified within the original model of Goldschmidt (which considers only the stretching and 
octahedral limits), within three other empirical models reported in Refs.~\citenum{Li2004, 
Giaquinta1994, Navrotsky1998}, and within our present model (see Figure~S7). Figure S7e shows that 
the stretching 
and octahedral limits correctly categorize nearly all perovskites in Database S1.1, but fail to 
discriminate against more than half of non-perovskites in Database S1.2. The empirical regions in 
Figure S7 b-d clearly demonstrate that by setting a lower bound to the tolerance factor, the 
accuracy of the model improves significantly for both perovskites and non-perovskites. However, up 
to now, this bound has been described via empirical data fitting. Our model predicts the tilt limit 
from first principles, all the while retaining a very good accuracy in distinguishing perovskites 
from non-perovskites. By comparison, the perovskite regions reported in Refs.~\citenum{Li2004, 
Giaquinta1994, Navrotsky1998} can be understood as zeroth- and first-order approximations to our
bounds, respectively.

By applying our classification algorithm to all possible 3,658,527 quaternary combinations, we
generated a library of 94,232 new perovskites and double perovskites that are expected to form with
a probability of 80\% (Figure~S8). The complete library of predicted perovskites is provided as
database~S2. Our library of future perovskites dwarves the set of all perovskites currently known,
and is comparable in size to the ICSD database of all known inorganic crystals, which contains
approximately 188,000 structures\cite{icsd}.

How many of our predicted perovskites are genuinely new compounds, i.e.\ have never been
synthesized? In order to answer this question we performed a large-scale web data extraction
operation by querying an internet search engine about each and every one of the nearly one hundred thousand
compounds in database~S2 (see Supplementary Information). This procedure revealed that the
overwhelming majority of these compounds have never been reported or mentioned before (database
S2.1), and that fewer than 1\% of the structures were already known, namely 786 out of 94,232
compounds (database~S2.2).

The 786 previously-known compounds reported in database~S2.2 were not included in our 
initial database~S1 of known crystals. We use this additional set of known compounds to perform a 
second blind test of our predictions. According to our inferential analysis we expect 
626 compounds of database S2.2 (79.7\% of 786) to be perovskites. By 
carrying out a manual literature search we confirmed that 555 crystals are indeed 
perovskites (database~S2.2.1). This result is remarkably consistent with our prediction. This blind 
test replaces a validation based on resource-intensive experimental synthesis of hundreds of new 
compounds with faster and inexpensive data analytics. The success of the blind test clearly 
demonstrates that, in spite of its simplicity, Goldschmidt's principle has a considerable predictive 
power.  Naturally, by combining our structure map with experiments and {\it ab initio} calculations 
on selected sub-families, the predictive accuracy of the model is bound to improve even further.

\begin{figure*}[t!]
\linespread{1.2}
\centering
\includegraphics[width=0.56\textwidth]{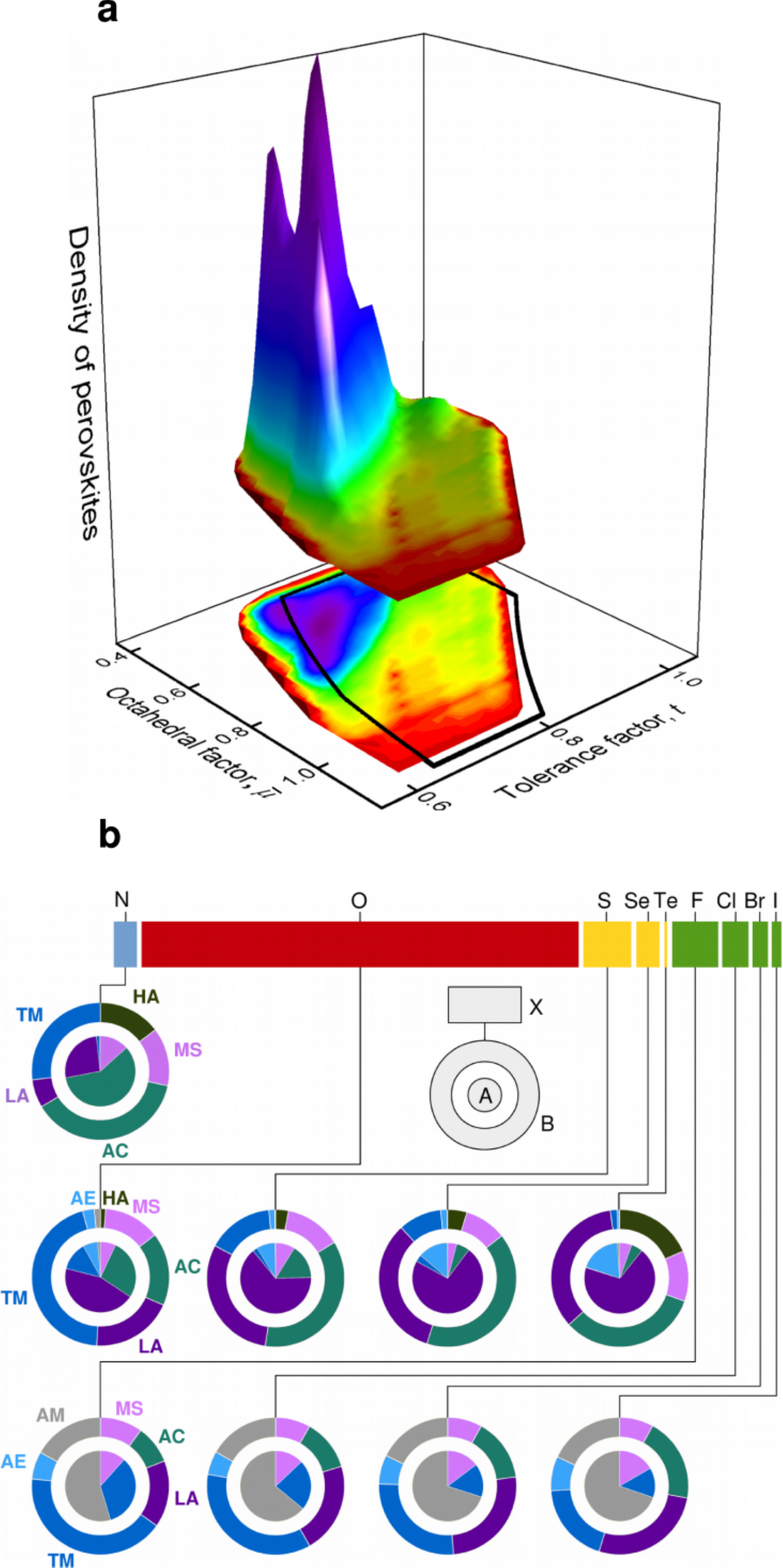}
\caption{{\bf Topography of the perovskite landscape}.
  {\bf a}, Density of predicted ternary an quaternary perovskites from databases~S1.1,
  S1.3 and S2. For clarity the density of perovskites has been integrated over all
  possible values of the octahedral mismatch $\Delta\mu$, so as to obtain a
  two-dimensional map in the $(\bar{\mu},t)$ plane. The two plots show the same quantity
  as a 2D colormap and a 3D surface, respectively. {\bf b}, Crystallographic site
  preference in databases~S1.1,~S1.3 and~S2. The horizontal bar illustrates the relative
  abundance of perovskites with a given anion X. For each anion, the rings illustrate the
  relative abundances of the A-site cation (inner ring) and of the B-site cation (outer
  ring). Cations are grouped using the standard classification: alkali metals (AM),
  alkaline earth metals (AE), transition metals (TM), lanthanides (LA), actinides (AC),
  metals and semiconductors (MS: Al, Ga, In, Sn, Tl, Pb, Bi, B, Si, Ge, As, Sb, Te, Po),
  and halogens (HA). Database~S2 contains 59 binary, 2834 ternary and 90,606 quaternary
  perovskites.
\label{fig3}}
\end{figure*}

What is the topography of our geometric structure map? Figure~\ref{fig3}a and Figure.~S9b show that
the majority of predicted perovskites tend to cluster towards the region with the lowest octahedral,
tolerance, and mismatch factors. This high density of compounds stems from the occurrence of a large
number of lanthanide oxide and actinide oxide perovskites, which tend to have similar geometric
descriptors due to the lanthanide contraction\cite{Pyykko}. We also note that the concentration of
compounds near the bottom of the map shows that the geometric tilt limit derived in this work
(Figure~\ref{fig1}d) is essential to accurately predict the formability of perovskites.

Figure~\ref{fig3}b shows the relative abundances of predicted perovskites. The majority of compounds 
are oxides (68\%), followed by halides (16\%), chalcogenides (12\%), and  nitrides (4\%). Why is the 
perovskite landscape dominated by oxides, and why nitrides are  so rare instead? To answer this 
question we observe that the -2 oxidation state of O admits as many as ten inequivalent 
charge-neutral combinations of the oxidation states of  the cations (Table~S2). Furthemore, the 
oxygen anion has a small radius (1.3~\AA), which is compatible with most transition metals, 
lanthanides, and actinides; and these elements form the most numerous groups in the Periodic Table. 
A similar argument could be made for chalcogens, which share the same oxidation state as oxygen. 
However, the chalcogen radii are too large (1.8-2.2 \AA) to accommodate most transition metals and 
actinides, hence chalcogenide perovskites constitute a much smaller family. Our finding is 
consistent with recent {\it ab initio} calculations\cite{Korbel}. Halide perovskites are even less 
numerous than chalcogenides, mostly owing to their more restrictive -1 oxidation number: in fact 
this oxidation state only admits +1 A-site cations and +1/+3 or +2/+2 B-site cations (Table~S2). 
Nitrides constitute an interesting exception to these trends. Indeed, while the ionic radius of N in 
the -3 oxidation state (1.5~\AA) is very similar to that of O and its oxidation number admits as 
many as seven inequivalent combinations of oxidation states for the cations, in all such 
combinations at least one B-site cation must have the unusually high +5 oxidation state (Table~S2). 
As the ionic radii tend to decrease with the oxidation number\cite{Pauling}, most B-site cations 
turn out to be too small to be coordinated by six nitrogen anions in an octahedral environment. As a 
result, if we exclude radioactive elements, we find fewer than eighty nitride perovskites across the
entire Periodic Table.  
We note that, owing to the lower electronegativity of nitrogen, the ionic character of 
the chemical bonds in nitride perovskites is reduced, and our geometric model is reaching its limits 
of applicability~\cite{Giaquinta1994}. However, the scarcity of nitride perovskites predicted by our 
model is fully consistent with recent {\it ab initio} calculations\cite{Sarmiento2016} and with 
experimental observations (only one ternary nitride perovskite can be found in ICSD), indicating 
that the rigid sphere approximation can still provide meaningful predictions for nitrides.
Among our predicted compounds we also identified many unexpected binary compounds of the 
type A$_2$X$_3$. One such example is iron oxide, Fe$_2$O$_3$. While this oxide is 
primarily known in the form of hematite (corundum structure), it was recently found 
that the crystal undergoes a phase transition to a perovskite at high pressure and 
temperature\cite{Bykova2016}. The stabilization of Fe$_2$O$_3$ as a perovskite under 
high pressure is in agreement with our  {\it ab initio} calculations (see discussion and
Figure~S10 of the Supporting Information) and can be associated with the well known phase transition 
of ilmenites (ternary ordered corundum) into perovskites, observed, for example, for 
FeTiO$_3$\cite{Navrotsky1998}. This finding suggests that several other binary compounds may hide a 
perovskite phase in their phase diagram, an intringuing possibility that is open to investigation.

\begin{figure*}[t!]
\linespread{1.2}
\centering
\includegraphics[width=0.7\textwidth]{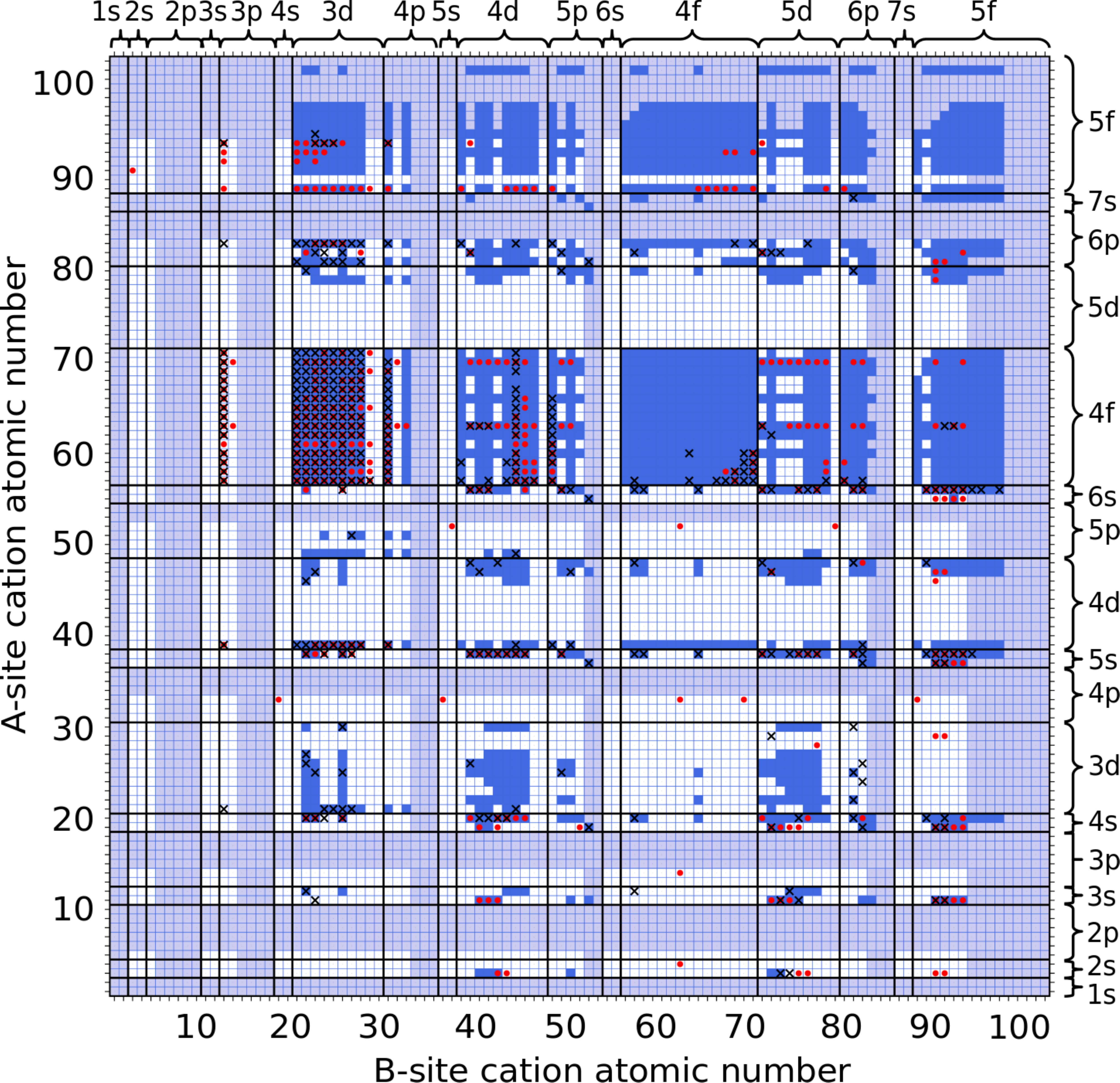}
\caption{{\bf Combinatorial screening of ternary oxide perovskites}. 
ABO$_3$ ternary compounds classified as perovskites by our geometric model (blue squares), {\it ab 
initio} calculations within the generalized gradient approximation to density functional theory, 
DFT/GGA (red disks) reported by Emery et al~\cite{Emery2016} and experimental data from Databases 
S1.1, S1.3 and S2.2.1 (black crosses). The gray shading highlights the region of the A/B map that is 
not taken into account in the {\it ab initio} screening. The curly bracket marks the empty valence 
shells corresponding to an interval of atomic numbers for A and B, e.g. the $3d$ interval 
corresponds to Z = 21-30 (Sn-Zn). Emery et al reported the study of the thermodynamic stability of 
5329 candidate ternary compounds, and found that 382 compositions are stable perovskites. There are 
383 ternary oxides in Databases S1.1, S1.3 and S2.2.1, DFT/GGA calculations predict 225 (59\%) of 
them to be stable while our model correctly classifies 354 of them (92\%). 
\label{fig4}}
\end{figure*}

To demonstrate the applicability of our model, we take the example of ternary oxide perovskites. 
 Figure~\ref{fig4} shows a comparison between the combinatorial screening of ternary oxides performed 
using our model, {\it ab initio} calculations reported in Ref. 37 and experimental data collected from 
Databases S1.1, S1.3 and S2.2.1. The compositions classified as perovskites by our model include 
92\% of the experimentally observed oxide perovskites. In particular, when A is a lanthanide and B 
is a first row transition metal, our model predicts that most compositions can form as a perovskite, 
in excellent agreement with DFT predictions and experiment. This can be explained by the similar 
ionic sizes of transition metals and rare earths respectively. The same prediction is made for the 
case when both A and B are rare earths, however fewer perovskites are found from DFT and experiment. 
The reason for this discrepancy is that the non-rattling principle effectively probes the dynamical
stability of a given chemical composition in the perovskite structure. However, it does not contain
information on its stability against decomposition (thermodynamic stability). Of the
two criteria, the thermodynamic stability requirement is more stringent, and this explains why
generally geometric blueprints tend to predict more perovskites than have actually been made or that
are predicted from {\it ab initio} calculations. Therefore, further theoretical and experimental
studies are required to ascertain whether these proposed compositions are also thermodynamically
stable. Despite this limitation, Figure~\ref{fig4} demonstrates that the Goldschmidt principle can be 
used as an efficient and reliable pre-screening tool for the high-throughput combinatorial design of 
perovskites. In fact, in 
Figure~\ref{fig4} we show that our model can reduce the  number of calculations by more than 70\% 
in the combinatorial screening of ternary oxides. Importantly, the Golschmidt no-rattling principle 
becomes increasingly useful in the context of screening all possible perovskites beyond oxides, 
reducing the total number of 3.6 million possible compositions by 97\%, to fewer than one hundred 
thousand candidates. Therefore, by leveraging the complementary strengths of Goldschmidt's empirical 
no-rattling principle and {\it ab initio} computational modelling it will be possible to explore the 
complete chemical landscape of all possible perovskites.

\section*{Conclusion}
In conclusion, we charted the complete landscape of all existing and future perovskites. By 
combining inferential statistics with large-scale web data extraction, we validated Goldschmidt's 
no-rattling principle on quantitative grounds, and developed a structure map to predict the 
stability of perovskites with a fidelity of 80\%. Our model completes the general 
theory that Goldschmidt proposed almost a century ago, and formalizes the non-rattling hypothesis 
into a mathematically rigurous set of criteria that can be used in the design and discovery of new 
perovskites.  As an outcome of our study, we were able to generate a library of almost one hundred 
thousand hitherto-unknown perovskites awaiting discovery (database~S2.1). By releasing this library 
in full, we hope that this work will stimulate much future experimental and computational research 
on these fascinating crystals. More generally, our findings suggest that geometric blueprints could
serve as a powerful tool to help tackle the exponential complexity of combinatorial materials design.

\vspace{0.5cm} 
\noindent
{\bf Methods.}
A full description of the methods, data provenance and statistical analysis used in 
this manuscript can be found in the Supporting Information.

\noindent
{\bf Acknowledgements.} This work was supported by the Leverhulme Trust (Grant No. RL-2012-001), the 
Graphene Flagship (Horizon 2020 Grant No. 696656-GrapheneCorel), and the UK Engineering 
and Physical Sciences Research Council (Grant No. EP/M020517/1).

\afterpage{\clearpage}
\afterpage{\clearpage}
\afterpage{\clearpage}
\afterpage{\clearpage}

\afterpage{\clearpage}

\newpage
\appendix 
\begin{widetext}
\section*{Supplementary Information}

\setlength{\parindent}{0pt}
{\bf Databases}

Data analyzed in this manuscript is reported in two main databases (available on 
ur group repository~\cite{github}), which are structured
as follows:
\begin{align*}
  \text{Database S1}       & = \begin{cases}
      \text{Database S1.1} & \text{Known perovskites.}
\\[0.15cm]
      \text{Database S1.2} & \text{Known non-perovskites.}
\\[0.15cm]
      \text{Database S1.3} & \text{Known compounds that can be both perovskites and}
\\[-0.15cm]
                           & \text{non-perovskites.}
    \end{cases} \\
\\
  \text{Database S2} & = \begin{cases}
      \text{Database S2.1} & \text{Predicted compounds that have never been made or}
\\[-0.15cm] 
                           &  \text{mentioned before.}
\\[0.15cm]
      \text{Database S2.2} & 
           \begin{cases}
             \text{Database S2.2.1} & \text{Predicted compounds that were found} 
\\[-0.15cm]
                                    & \text{on the internet and are perovskites.} 
\\[0.15cm]
             \text{Database S2.2.2} & \text{Predicted compounds that were found}
\\[-0.15cm]
                                    & \text{on the internet and are not perovskites.} 
           \end{cases}
    \end{cases}
\\[0.15cm]
\end{align*}
\end{widetext}
\setlength{\parindent}{0pt}
%

{\bf Data provenance}

The database~S1 of known compounds is generated by collecting structures from the Inorganic Crystal
Structure Database (ICSD)\cite{icsd} and from Refs.~\citenum{Li2004,Li2008, Galasso, Vasala2015, 
Flerov1998, Meyer1982, Flerov2001}. From Ref.~\citenum{Vasala2015} we consider only quaternary 
compounds. The calculation of $(\bar{\mu},t,\Delta\mu)$ is performed by using the oxidations states 
reported in these references. For ternary and quaternary perovskites we search the ICSD by structure 
type, using the keywords `perovskite', `elpasolite', or `LiNbO$_3$'. Out of the compounds returned 
by these searches we retain only pure crystals, i.e.\ compounds with integer site occupation numbers 
and integer oxidation states. We discard theoretically-predicted compounds that have not been 
synthesized.

In order to find ternary non-perovskites in the ICSD, we search both by structure type and by ANX 
formula, following the classification of Refs.~\citenum{Muller-Roy,Meyer1982}. We consider the 
following structure types: spinel, ilmenite, aragonite, calcite, olivine, pyroxene, BaNiO$_3$, 
BaRuO$_3$, BaMnO$_3$; and the following ANX formulae: A$_2$BX$_4$, A$_3$BX$_5$, AB$_2$X$_5$, 
AB$_3$X$_7$, A$_2$B$_3$X$_8$, AB$_4$X$_9$ and AB$_3$X$_8$. We retain the compounds which fulfil the 
charge neutrality rule and Pauling's valency rule for an ABX$_3$ stoichiometry.

In order to collect quaternary non-perovskites from the ICSD, we first search for quaternary 
non-perovskites which have the same general formula A$_2$BB$'$X$_6$ as for double perovskites. For 
the other possible stoichiometric compositions we could not find an established structural 
classification as for ternary non-perovskites. To overcome this lack of information, we considered 
crystals with formulas A$_m$B$_n$B$'_p$X$_q$ which fulfil the charge neutrality and valency 
conditions, $2a+b+b'-6x=0$ and $a\leq (b+b')/2$, where $a$, $b$, $b'$, and $x$ are the oxidation 
numbers of A, B, B$'$, and X, respectively. This linear system admits infinite solutions, therefore 
we limit the search to the cases $q=5,6,7$, so as to span the neighborhood of $q=6$ in 
A$_2$BB$'$X$_6$ compounds.  This choice corresponds to searching the ICSD using the following ANX 
formulae:
ABB$'$X$_5$, 
AB$_2$B$'$X$_6$,
AB$_2$B$'$X$_7$, 
AB$_3$B$'$X$_6$,
AB$_3$B$'$X$_7$,
AB$_4$B$'$X$_6$,
AB$_5$B$'$X$_7$,
AB$_6$B$'$X$_7$.
A$_3$BB$'$X$_7$,
A$_2$B$_3$B$'$X$_7$,~and~A$_4$BB$'$X$_7$.

After collecting all data we remove duplicates. In addition, we cross-check the lists of perovskites 
and non-perovskite to identify those compositions for which both perovskite and non-perovskite 
compounds are found.  We find that 77 compounds (database~S1.3) crystallize either as a perovskite
or a non-perovskite.

The compounds obtained from these searches form the set of previously-known compounds, and are 
reported in database~S1, together with the source references and the reported oxidation states. The 
database contains 77 compounds that can be either perovskites or another structure (database~S1.3), 
345 ABX$_3$ perovskites and 1277 A$_2$BB$'$X$_6$ double perovskites (database S1.1), and 592 
compounds that are neither a perovskite nor a double perovskite (database S1.2).\\

{\bf Geometric bounds resulting from the no-rattling principle: ABX$_3$ perovskites}

As in the main text we indicate by $r_{\rm A}$, $r_{\rm B}$, and $r_{\rm X}$ the ionic radii of A, 
B, and X, respectively. From these values we obtain the tolerance factor as\cite{Goldschmidt}:
 \begin{equation}\label{eq.t}
  t = (r_{\rm A}+r_{\rm X})/\sqrt{2}(r_{\rm B}+r_{\rm X}),
  \end{equation}
and the octahedral factor as:
  \begin{equation}\label{eq.mu}
   \mu = r_{\rm B}/r_{\rm X}.
  \end{equation} 
Following Goldschmidt's principle we consider that in ABX$_3$ perovskites the B cation and the 
X anion are always in contact. Instead the A cation may or may not touch other anions or cations. 
The geometric bounds defining the stability region correspond to extremal packing configurations 
whereby A-A$^*$, A-B, A-X, or X-X$^*$ touch, having indicated by an asterisc a 
nearest-neighbor ion of the same type.  In the following we consider the Platonic model of 
perovskites\cite{Filip2014}, whereby the BX$_6$ octahedra are ideal and corner-sharing. 
We discuss the geometric limits using the naming convention of Figure~1: stretch limit (SL), 
octahedral limit (OL), tilt limits (TL1 and TL2), chemical limits (CL1 and CL2), and secondary 
stretch limits (SSL1 and SSL2). 

\underline{Stretch limit (SL)}

The stretch limit corresponds to the situation where the A cation is so large that it touches all 
twelve X anions defining the octahedral cavity; this can only happen in a cubic perovskite, as 
shown in Figure~1b. In this configuration the distances between the centers of A and X and between 
the centers of B and X are related by $\sqrt{2}(r_{\rm B}+r_{\rm X}) = r_{\rm A}+r_{\rm X}$. After 
dividing by the left-hand side we obtain $t=1$, irrespective of $\mu$. In Figure~1e this boundary is 
labelled `SL'.

\underline{Octahedral limit (OL)}

The octahedral limit corresponds to the situation where two X anions belonging to the same BX$_6$ 
octahedron are in contact, as shown in Figure~1c. In this configuration the distance between the 
centers of the anions satisfies the condition $\sqrt{2}(r_{\rm B}+r_{\rm X}) = 2r_{\rm X}$. After 
dividing both sides by $r_{\rm X}$ we find $\mu = \sqrt{2}-1$, irrespective of $t$. This bound is 
labelled `OL' in Figure~1e.

\underline{Tilt limits (TL1 and TL2)}

In ABX$_3$ perovskites, adjacent BX$_6$ octahedra can tilt until nearest-neighbor anions 
from each octahedron make contact. In this configuration the octahedral cavity is distorted
from a regular cuboctahedron; to satisfy Goldschmidt's no-rattling principle the A
cation must move off-center, as shown in Figure~1e.\\ In order to investigate this extremal
configuration we employ the Platonic model of perovskites developed in Ref.~\citenum{Filip2014}.
This model considers a orthorhombic perovskite unit cell consisting of four BX$_6$ octahedra,
belonging to the $Pnam$ space group; all atomic coordinates are uniquely defined by
specifying the B-X bond length, $r_{\rm B}+r_{\rm X}$, and two angles, the tilt angle $\theta$ 
of an octahedron with respect to the $z$ axis, and the precession $\phi$ of this octahedron 
around the same axis. The coordinate system is shown in Figure~S2. The Cartesian 
coordinates of the anions indicated in Figure~S2 can be expressed as follows:
  \begin{eqnarray}
   {\rm X}_1   &: &\phantom{-}a(-x_{\rm e}+1/2);  \phantom{-}b(\phantom{-}y_{\rm e}+1/2); 
             -c(\phantom{-}z_{\rm e}\phantom{+1/22}), \label{eq.1}\\
   {\rm X}_2   &: &\phantom{-}a(-x_{\rm e}+1/2);  \phantom{-}b(\phantom{-}y_{\rm e}+1/2); 
      \phantom{-}c(\phantom{-}z_{\rm e}+1/2), \\
   {\rm X}_3   &: &\phantom{-}a(\phantom{-}x_{\rm e}-1/2);  \phantom{-}b(-y_{\rm e}+1/2); 
      \phantom{-}c(-z_{\rm e}+1/2), \\
   {\rm X}_4   &: &-a(\phantom{-}x_{\rm e}\phantom{-1/21});  
    \phantom{-}b(-y_{\rm e}+1\phantom{/2}); 
                        -c(\phantom{-}z_{\rm e}\phantom{-1/21}), \\
   {\rm X}_5   &: &-a(\phantom{-}x_{\rm e}\phantom{-1/21});  
    \phantom{-}b(-y_{\rm e}+1\phantom{/2}); 
                        \phantom{-}c(\phantom{-}z_{\rm e}+1/2), \\
   {\rm X}_6&: & \phantom{-}a(\phantom{-}x_{\rm a}\phantom{-1/21});
     \phantom{-}b(\phantom{-}y_{\rm a}\phantom{+1/11});
            \phantom{-}c(\phantom{-}z_{\rm a}\phantom{+1/11}), \\
   {\rm X}_7&: &\phantom{-}a(\phantom{-}x_{\rm a}-1/2);  \phantom{-}b(-y_{\rm a}+1/2);
             \phantom{-}c(\phantom{-}z_{\rm a}\phantom{+1/21}). \label{eq.7}
  \end{eqnarray} 
The fractional coordinates and the lattice parameters appearing in these expressions
are given by:
\begin{widetext}
  \begin{eqnarray}
    && x_{\rm e} =  \frac{1}{4}(1+\cos{\theta}\tan{\phi}); \qquad
       y_{\rm e}  =  \frac{1}{4}(1-\tan{\phi}/\cos{\theta}); \qquad
       z_{\rm e}  =  -\frac{\sqrt{2}}{8}\tan{\theta}, \label{eq.7}\\ 
    && x_{\rm a}  =   \frac{\sqrt{2}}{4} \tan{\phi}\sin{\theta}; \hspace{1.45cm}
       y_{\rm a}  =  \frac{\sqrt{2}}{4} \tan{\theta}; \hspace{2.6cm}
       z_{\rm a}  =  \frac{1}{4}, \\
    && a = 2\sqrt{2}(r_{\rm B}+r_{\rm X})\cos{\phi}; \hspace{0.7cm}
       b = 2\sqrt{2}(r_{\rm B}+r_{\rm X})\cos{\theta}\cos{\phi}; \hspace{0.1cm}
       c = 4(r_{\rm B}+r_{\rm X})\cos{\theta},\label{eq.9}
  \end{eqnarray}
\end{widetext}
where we have obtained $x_{\rm e}$, $y_{\rm e}$, $z_{\rm e}$, $x_{\rm a}$, 
$y_{\rm a}$ and $z_{\rm a}$ following the procedure described in Ref.~\citenum{Filip2014},
but using the transpose of the rotation matrix.
Upon tilting the octahedra, the distance between the anion pairs X$_1$-X$_2$, X$_2$-X$_3$, 
X$_4$-X$_5$, and X$_2$-X$_6$ in Figure~S2 decreases. We can ignore the pair X$_4$-X$_5$ which is
always found at the same distance as X$_1$-X$_2$. In this case there are two mutally-exclusive
extremal configurations: (1) X$_1$ and X$_2$ as well as X$_2$ and X$_3$ are in contact; (2) X$_1$ 
and X$_2$ as well as X$_2$ and X$_6$ are in contact. In both cases the anions X$_1$ and X$_2$ touch, 
so we examine this condition first. By requiring that the distance between the centers of X$_1$ and 
X$_2$ is $2r_{\rm X}$ we can determine the tilt angle $\theta$ in either extremal configuration:
  \begin{equation}\label{eq.10}
  \cos\theta = (2p + \sqrt{3-2p^2})/3, \,\,\, \mbox{ with }\,\,\, p = 1/(\mu+1).
  \end{equation}
By further requiring that the distance between X$_2$ and X$_3$ or that the distance between X$_2$ 
and X$_6$ is equal to $2r_{\rm X}$, we obtain the two extremal precession angles $\phi_1$ and 
$\phi_2$: 
  \begin{eqnarray}\label{phmax}
  \sin 2\phi_1  &=& (1-p^2)/\cos\theta, \\
  \tan \phi_2  &=& (\cos\theta-s)/(1-2p^2), 
  \end{eqnarray}
having defined:
  \begin{equation}
    s^2=\cos^2{\theta}-(1-2p^2)(2\cos^2\theta-\sqrt{2}\sin 2\theta +1-2p^2). \label{eq.13}
  \end{equation}
These conditions allow us to determine, for each octahedral factor $\mu$, the extremal 
configurations corresponding to the maximum possible octahedral tilt. In order to obtain a boundary 
for the stability region, we need to find the limit of the tolerance factor $t$ in these 
configurations. To this aim we determine the size of the largest A cation which fits in the cavity 
and touches as many X anions as possible.\\
By symmetry the center of this A cation along the $c$ axis must be $z_{\rm A} = 1/4$. 
To find the other coordinates $x_{\rm A}$ and $y_{\rm A}$, we require that they allow us to fit
the largest possible A cation in the cavity. This statement corresponds to asking that 
$(x_{\rm A},y_{\rm A},z_{\rm A})$ be the center of a sphere passing through the centers of
X$_1$, X$_2$, X$_4$, X$_5$ and X$_7$. 
Using Eqs.~(\ref{eq.1})-(\ref{eq.9}) we obtain the following conditions for $x_{\rm A}$ and 
$y_{\rm A}$:
\begin{widetext}
  \begin{eqnarray}
  2a^2x_{\rm A}(1-x_{\rm a}-x_{\rm e}) 
  + 2b^2 y_{\rm A}(y_{\rm a}+y_{\rm e})&=&a^2(x_{\rm e}^2-x_{\rm a}^2+x_{\rm a}- x_{\rm e})+  \\ \nonumber
  &+&b^2(y_{\rm e}^2-y_{\rm a}^2 + y_{\rm a}+y_{\rm e}) +(1/4+z_{\rm e})^2c^2,\label{eq.14} \\
  a^2x_{\rm A}+b^2y_{\rm A}(4y_{\rm e}-1)  &=&a^2(1/4-x_{\rm e})+b^2(3y_{\rm e}-3/4).  \label{eq.15}
  \end{eqnarray}
\end{widetext}
Furthermore the radius of the sphere is given by:
  \begin{equation}\label{eq.16}
  r_\mu = [(x_{\rm A}-x_{\rm a}+1/2)^2a^2+(y_{\rm A}+y_{\rm a}-1/2)^2b^2]^{1/2}.
  \end{equation}
In order to  fulfil the no-rattling principle, the A cation must touch the nearest neighbor X anions.
This condition implies that the radius of the sphere must equal the distance between A and X,
$r_\mu =  r_{\rm A}+r_{\rm X}$. As a result we find that the boundary for the tolerance factor is
  \begin{equation}
  t = r_\mu/\sqrt{2}(r_{\rm B}+r_{\rm X}).\label{eq10}
  \end{equation}
To proceed we evaluate this boundary numerically as follows: we consider a range of octahedral 
ratios $\mu$; for each value of $\mu$ we calculate the angles $\theta$ and $\phi_1$, $\phi_2$ 
through Eqs.~(\ref{eq.10})-(\ref{eq.13}). Starting from these angles we obtain $x_{\rm e}$, 
$y_{\rm e}$, $z_{\rm e}$ and $x_{\rm a}$, $y_{\rm a}$, $z_{\rm a}$ via Eqs.~(\ref{eq.7})-(\ref{eq.9}). 
We then solve Eqs.~(17)-(18) to determine $x_{\rm A}$ and $y_{\rm A}$. 
Finally we use these results inside Eq.~(\ref{eq.16}) to find~$r_\mu$.\\
A plot of $\rho_\mu=r_\mu/r_{\rm X}$ vs.\ $\mu$ is shown in Figure~S3. The two extremal 
configurations define two approximately straight lines; instead of repeating the above steps for 
every value of $\mu$ used in the structure maps of the main text, we fit the lines using a 
piece-wise linear function:
\begin{widetext}
  \begin{equation}\label{eq11}
  \rho_\mu = 1.366+0.442\bar{\mu}\,\mbox{ for }\,\bar{\mu} < 0.8,\qquad
  \rho_\mu = 1.125+0.732\bar{\mu}\,\mbox{ for }\,\bar{\mu}> 0.8.
  \end{equation}
\end{widetext}
Using these linear fits inside Eq.~(\ref{eq10}) yields the tilt limits TL1 and TL2 shown in Figure~1e.  

\underline{Secondary stretch limits (SSL1 and SSL2)}

An additional geometric limit is obtained by considering the extremal configuration wherein two 
adjacent A-site cations are in contact. The configuration with the largest possible such cations is 
cubic, as shown in Figure~S4a.  In this configuration the distance between the centers of A and 
A$^*$, $2(r_{\rm B}+r_{\rm X})$, equals the diameter of each cation, $2r_{\rm A}$. From this 
equality, by using the relations between $\mu$, $t$ and $r_{\rm A}/r_{\rm X}$, $r_{\rm B}/r_{\rm X}$ 
we obtain the boundary $t = (\mu+2)/\sqrt{2}(\mu+1)$.  This limit is indicated as SSL1 in Figure~1e. 
Similarly it is straightforward to verify that in the extremal configuration wherein the A and B 
cations are in contact one has $t = \sqrt{3}/\sqrt{2}-(\mu-1)/\sqrt{2}(\mu+1)$ (Figure~S4b). This 
boundary is labelled SSL2 in Figure~1e. Since the boundaries SSL1 and SSL2 lie above the boundary SL 
in Figure~1e, the upper limit of the tolerance factor $t$ is set by the stretch limit.

\underline{Chemical limits (CL1 and CL2)}

The largest cation and the smallest anion in the Periodic Table, Cs and F respectively, set an upper 
bound for the ratio $r_{\rm A}/r_{\rm X}$. Since by definition we have 
$r_{\rm A}/r_{\rm X} = \sqrt{2}t(\mu+1)-1$, there will not be any compounds in the perovskite map 
above the line $t=(r_{\rm Cs}/r_{\rm F}+1)/\sqrt{2}(\mu+1)$. This boundary is indicated as CL1 in 
Figure~1e. A similar reasoning applies to the octahedral ratio 
$\bar{\mu} = (r_{\rm B}+r_{{\rm B}'})/2r_{\rm X}$. Among all perovskites and double perovskites in 
database~S1 this ratio is largest for the combination of Fr$^+$ and Ac$^{3+}$ at the B and B$'$ 
sites, respectively, and F at the X site ($\bar{\mu}=1.136$). In Figure~1e this boundary is labelled 
CL2.\\

{\bf Geometric bounds resulting from the no-rattling principle: A$_2$BB$'$X$_6$ double perovskites}

In order to derive the geometric bounds for quaternary A$_2$BB$'$X$_6$ double perovskites, in this 
work we introduce generalized Goldschmidt's parameters as follows: 
(i) the average octahedral factor:
 \begin{equation}\label{eq.mu4}
     \bar{\mu} = (r_{\rm B}+r_{{\rm B}'})/2r_{\rm X},
 \end{equation}
(ii) the octahedral mismatch:
 \begin{equation}\label{eq.dmu4}
   \Delta \mu = |r_{\rm B}-r_{{\rm B}'}|/2r_{\rm X},
 \end{equation}
and (iii) the generalized tolerance factor:
 \begin{equation}\label{eq.t4}
    t = \frac{r_{\rm A}+r_{\rm X}}{\sqrt{2} 
   \{[(r_{\rm B}+r_{{\rm B}'})/2+r_{\rm X}]^2+ (r_{\rm B}-r_{{\rm B}'})^2/4\}^{1/2}}.
 \end{equation}
It is immediate to verify that when $r_{\rm B}=r_{{\rm B}'}$ we obtain $\Delta \mu=0$,
and the tolerance and octahedral factors in Eqs.~(\ref{eq.t4}) and (\ref{eq.mu4}) 
reduce to the standard definitions used for ternary perovskites in Eqs.~(\ref{eq.t}) and 
(\ref{eq.mu}).

\underline{Stretch limit}

The stretch limit corresponds to the extremal configuration wherein the A cation is in contact with 
the X anions. At variance with the case of ternary perovskites, in quaternary systems the X anion is 
no longer located midway between the B and B$'$ cations, as shown in Figure~S1b. By applying 
Pitagoras' theorem to the triangle in Figure~S1c we find: 
\begin{widetext}
 \begin{equation}
 (r_{\rm A}+r_{\rm X})^2 = 2\left( \frac{r_{\rm B}+2r_{\rm X}+r_{{\rm B}'}}{2} \right)^2+
        \left[ (r_{\rm B}+r_{\rm X})- \frac{r_{\rm B}+2r_{\rm X}+r_{{\rm B}'}}{2} \right]^2.
 \end{equation}
\end{widetext}
After combining this equation with Eq.~(\ref{eq.mu4})-(\ref{eq.t4}) we obtain the boundary $t=1$,
irrespective of $\bar{\mu}$ and $\Delta\mu$, precisely as for ternary perovskites.

\underline{Octahedral limit}

The octahedral limit for double perovskites is found by considering the extremal configuration where
two X anions belonging to the smallest octahedron among BX$_6$ and B$'$X$_6$ are in contact. The 
smallest radius among $r_{\rm B}$ and $r_{{\rm B}'}$ is given by $r_{\rm min}=
(\bar{\mu}-\Delta\mu)r_{\rm X}$, therefore the octahedral limit for the corresponding octahedron 
reads $\sqrt{2}(r_{\rm min}+r_{\rm X}) = 2r_{\rm X}$, as in the case of ternary perovskites. After 
dividing by $r_{\rm X}$ we obtain the boundary plane $\Delta\mu = \bar{\mu}+\sqrt{2}-1$ 
(irrespective of $t$).  This boundary can be seen in Figure~2a as the surface of the stability 
region which forms an angle of 45$^\circ$ with the horizontal plane.

\underline{Tilt limit}

In order to derive the tilt limit it is necessary to generalize the Platonic model of 
Ref.~\citenum{Filip2014} to the case of double perovskites. However, since the majority of 
structures reported in database~S1 of known perovskites and in database~S2 of predicted perovskites 
have small octahedral mismatch ($\Delta\mu<0.2$ for 95\% of all compounds, Figure~S8), and since by 
construction the tilt limit for $\Delta\mu=0$ reduces to that of ternary perovskites, it is 
legitimate to use Eqs.~(\ref{eq10}) and (\ref{eq11}) also for quaternary perovskites. The small 
error involved in this approximation is absorbed in the optimal formability, as determined in our 
statistical analysis below.

\underline{Chemical limits}

The same limits CL1 and CL2 apply to ternary and quaternary perovskites.\\

{\bf Uncertainty quantification, optimum formability, and classification accuracy}

To each compound in database~S1 we associate a rectangular cuboid in the $(\bar{\mu}, t, \Delta\mu)$
space. The corners of the cuboids correspond to the minima and maxima of $t$, $\bar{\mu}$, and
$\Delta\mu$ for each compound, as calculated using available ionic radii\cite{Shannon, shannon-web}.
For B-site cations we consider the ionic radii corresponding to six-fold coordination; for A and X 
cations we consider all available coordination numbers, since the coordination of these ions is 
sensitive to the octahedral tilt. When both high-spin and low-spin radii are available, we consider 
the high-spin state.

For each compound we define the formability $f$ as the fraction of the cuboid volume falling within 
the stability region. To calculate the formability we discretize each cuboid using a 
5$\times$10$\times$5 mesh of points in the $(\bar{\mu},t,\Delta\mu)$ space, and count the points 
that satisfy the geometric criteria set in Table~S1. The largest formability is $f=1$ (the cuboid is 
entirely inside the stability region), the smallest formability is $f=0$ (cuboid outside of the 
region).  We classify a compound as a `perovskite' if the calculated formability is above a critical 
value $f_{\rm c}$ (to be determined), and as `non-perovskite' otherwise. 

  \begin{table*}[t!]
  \vspace{-0.7cm}
  \begin{center}
  \begin{tabular}{|l l l|}
  \hline
  {\bf Bound type} & {\bf Abbrev.}  & {\bf Inequality} \\
  \hline
  Stretch limit          & SL     & $t \leq 1$ \\
  \hline
  Octahedral limit &   OL   & $\bar{\mu}\geq \sqrt{2}-1 + \Delta \mu$ \\
  \hline
  Tilt limit & TL1 & $\displaystyle{ t \geq (0.44~\bar{\mu}+1.37)/\sqrt{2}(\bar{\mu}+1)}$  \\
  & TL2 & $\displaystyle{ t \geq (0.73~\bar{\mu}+1.13)/\sqrt{2}(\bar{\mu}+1)}$  \\
  \hline
  Chemical limit & CL1 &$\displaystyle{t \leq 2.46/[2(\bar{\mu}+1)^2+\Delta\mu^2}]^{1/2}$ \\
  & CL2 &$\bar{\mu} \leq 1.05$ \\
  \hline
  \end{tabular}\vspace{-5pt}
  \end{center}
  \linespread{1.1}
  \caption{{\bf Geometric limits for the formability of perovskites}. Summary of the 
  inequalities defining the stability region of perovskites and double perovskites according to 
  Goldschmidt's no-rattling principle (blue volume in Figure~2A). The definitions of the generalized 
  tolerance factor $t$, the generalized octahedral factor $\bar{\mu}$, and the octahedral mismatch 
  $\Delta\mu$ are given in Eqs.~(\ref{eq.mu4}) through (\ref{eq.t4}), respectively.\label{tb1}}
  
  \end{table*}

In order to determine the critical value $f_{\rm c}$ we proceed as follows. For each value of 
$f_{\rm c}$ we evaluate the classification accuracy of all the perovskites and of all the 
non-perovskites in database~S1. The classification accuracy is defined as the number of compounds 
correctly classified divided by the number of all compounds of a given type. The variation of the 
classification accuracy with $f_{\rm c}$ is shown in Figure~S6a. It is seen that the classification 
accuracy improves towards large $f_{\rm c}$ for perovskites, and towards small $f_{\rm c}$ for 
non-perovskites. The value at which the two curves intersect defines the critical formability, 
$f_{\rm c}=0.34$. Using this value as the dividing line between perovskites and non-perovskites, we 
achieve a classification accuracy of 79.7\% for both perovskites and non-perovskites.  Having the 
same classification accuracy for both groups is important in view of making predictions for datasets 
where it is not known which and how many compounds belong to either group.

  \begin{table}[h!]
  \begin{center}
  \begin{tabular}{| c   c   c   c   c |}
  \hline
  \phantom{$-$}{\bf X} & \,\,{\bf A}  &  \,\,{\bf B} &  \,\,{\bf B$'$} & {\bf Family} \\
  \hline
  $-1$ & $+1$	& $+1$  & $+3$  & Halides \\
  &    & $+2$  & $+2$  &                     \\
  \hline 
  $-2$ & $+1$ & $+3$  & $+7$ & Oxides and \\
  &    & $+4$  & $+6$ & chalcogenides\\ 
  &    & $+5$  & $+5$ &  \\
  & $+2$ & $+1$  & $+7$ &   \\
  &    & $+2$  & $+6$ &        \\
  &    & $+3$  & $+5$ &        \\ 
  &    & $+4$  & $+4$ &        \\
  & $+3$ & $+1$  & $+5$ &      \\ 
  &    & $+2$  & $+4$ &       \\ 
  &    & $+3$  & $+3$ &       \\
  \hline 
  $-3$ & $+2$ & $+7$  & $+7$ & Nitrides \\
  & $+3$ & $+6$  &$+6$ &            \\
  &    & $+5$  & $+7$ &            \\ 
  & $+4$ & $+3$  & $+7$ &           \\
  &    & $+4$  & $+6$ &              \\
  &    & $+5$  & $+5$ &           \\
  \hline
  \end{tabular}\vspace{-5pt}
  \end{center}
  \linespread{1.1}
  \caption{{\bf Cation oxidation states in perovskites and double perovskites.} For each 
  oxidation state of the anion X, the possible oxidation states of the cations A, B, and B$'$ are 
  constrained by the charge neutrality condition and Pauling's valency rule. The oxidation state of 
  oxygen and chalcogens admits the largest number of combinations of cation oxidation states. 
  \label{tb2}} 
  \end{table}

To estimate the statistical uncertainty we make use of the central limit theorem. From all compounds 
in database~S1.1 and S1.2, including both perovskites and non-perovskites, we randomly draw 50,000 
subsets of $N$ compounds (with $N$ between 5 and 400), and we evaluate the classification accuracy 
for each subset. The distribution of these values tends to a Gaussian, as shown in Figure~S6b.  The 
standard deviation of this distribution depends on the size $N$ of the subset, as shown in 
Figure~S6c. For sets of 100 compounds or more, the fidelity of our classification scheme is found to 
be 79.7$\pm$4.0\% with a confidence level of 95\%. As a result, if we consider sets of 100 ABX$_3$ 
or A$_2$BB$'$X$_6$ unkown compounds, in 95\% of cases our geometric algorithm is expected to 
correctly classify between 76 and 86 compounds.\\

{\bf High-throughput screening of all possible ABX$_3$ and A$_2$BB$'$X$_6$ compounds}

In order to identify all possible perovskites and double perovskites we proceed as shown 
schematically in Figure~S6. We first construct all possible combinations of A, B, B$'$ cations and 
X anions from the Periodic Table, by considering all elements with associated ionic radii in 
Shannon's database\cite{Shannon, shannon-web}. We obtain 3,658,527 hypothetical compounds. Based on 
electrostatic considerations we exclude the combinations that do not satisfy the charge neutrality 
condition, that is $a+b+3x=0$ where $a$, $b$, and $x$ denote the oxidation numbers of A, B, and X, 
respectively (and similarly for double perovskites). After this refinement we are left with 
1,439,753 hypothetical compounds, including duplicate structures which differ only by the oxidation 
states of their elements. We further refine this dataset by retaining only compounds that fulfil 
Pauling's valency rule, whereby the oxidation number of A does not exceed that of B\cite{Pauling}; 
after this refinement we are left with 1,131,737 structures. For each of these combinations we 
calculate its rectangular cuboid in the $(\bar{\mu},t,\Delta\mu)$ space and the associated 
formability $f$. We find 94,232 distinct compounds that pass the test $f>f_{\rm c}=0.34$ and which 
do not belong to the set of known perovskites in databases~S1.1 or S1.3. The final set of predicted 
perovskites and double perovskites is reported in database~S2.\\

{\bf Data mining the internet}

In order to verify that the 94,232 compounds of database~S2 have not hitherto been reported, we 
performed a high-throughput internet search of all their chemical formulas. We attempted to use 
Google, but automated searches are considered as web-scraping by this engine, and are not allowed. 
We therefore opted for DuckDuckGo (DDG), which can be used for high-throughput searches. DDG
generates results from many sources including Yahoo!, Wikipedia, and Bing\cite{wikiduck}. Using a 
Unix tcsh shell we executed the following search key for each compound:\vspace{-6pt}

\begin{widetext}
{\footnotesize
\begin{verbatim}
w3m "http://duckduckgo.com/?q=\${compound}" 
-dump | grep \${compound} | grep -v Search | grep -v 'Limit results to'
\end{verbatim}\vspace{-7pt}}
\end{widetext}

where `\$\{compound\}' is a plain string with the chemical formula of the compound, for example
`Cs2BiAgCl6'. The compound has been found on the internet if this shell command returns a non-empty 
string. To test the procedure we executed this command for all the known perovskites in 
database~S1.1, and S1.3 and we correctly found 1254 out of 1699 perovskites. Therefore the success 
rate of this data mining procedure is 74\%.

By executing the above command for the 94,232 compounds of database~S2 we found that 92,899 
compounds do not have any presence on the internet, and hence can be considered as new perovskites. 
1333 compound formulas were found to have a presence on the internet. We therefore investigated each 
of these compounds by performing manual internet searches. We retained only compounds which appeared 
in scientific documents (such as peer-reviewed publications, conference abstracts or proceedings, 
national laboratory reports or masters and doctoral dissertations) and which contained a minimal 
Vdiscussion of the crystal structure. Using this procedure we unambiguously identified 786 ternary 
and quaternary crystals, out of which 555 are perovskites. These 786 compounds were used for our 
blind test of the classification accuracy, and are reported in database~S2.2.1 (perovskites) and 
S2.2.2 (non-perovskites) alongside the literature references that we used to confirm their 
classification. After removing the 786 compounds found from internet mining, we obtain a set of 
93,447 predicted perovskites never hitherto reported, listed in database S2.1.

{\bf Prediction of Fe$_2$O$_3$ perovskite phase}

Experimental observations of the transition of Fe$_2$O$_3$ into the perovskite phase at high
pressure have been previously reported in Refs.~\citenum{Ono2004-1, Bykova2016}. This transition 
could be rationalized qualitatively by comparison with the ilmenite to perovskite phase transition 
which occurs, for example in  FeTiO$_3$\cite{Navrotsky1998}, and involves a rearrangement of the 
position of the cations types in the alternating layers, as shown in Figure~S10b-c. The corundum 
structure of Fe$_2$O$_3$ can be related to the ilmenite structure of FeTiO$_3$ simply by replacing 
the Ti with Fe, and it is therefore possible that a similar phase transition mechanism as that of 
FeTiO$_3$ occurs for Fe$_2$O$_3$ at high pressure.

We have performed a set of preliminary calculations to assess of the stability of Fe$_2$O$_3$ in a
perovskite crystal structure, with respect to its known stable structure at ambient temperature and
pressure condition, the corundum structure. In Figure~S10 we show a comparison of the enthalpies
calculated within DFT as a function of pressure between 0 and 200 GPa for the corundum and
perovskite phases of Fe$_2$O$_3$. We note that our calculations do not include the effect of
temperature and zero-point motion. As shown in Figure~S10, at 0 GPa the corundum structure is more
stable than the perovskite structure by approximately 20~kJ/mol, but the formation enthalpies of the
two structures become closer as the pressure increases. In the inset of Figure~S10 we show that at
pressures beyond 140 GPa the perovskite phase becomes more stable than the corundum phase, in
agreement with the predictions of our model. We note that transition pressure of 140 GPa is
significantly overestimated with respect to the experimental measurement of 30-45 GPa reported in
Refs.~\cite{Ono2004-1, Bykova2016}. To achieve a better agreement with experiment, more refined 
calculations of the phase diagram of Fe-O are required, which should include a comparison between 
all possible competing phases of Fe$_2$O$_3$,the effect of  vibrational entropy, as well as a thorough 
assessment of the total energy  for different exchange correlation functionals and different magnetic 
configurations of Fe$_2$O$_3$.

From our preliminary DFT calculations we estimate that the average ionic size for the Fe$_2$O$_3$, 
$\displaystyle{R_0 = \Big(3\Omega/4\pi\Big)^{1/3}}$\, where $\Omega$ is the unit cell volume,
increases by 12\% when with pressure increasing from 0 to 200~GPa. On the other hand, the ionic radius 
of Fe$^{3+}$ takes values between 0.49~\AA\ (4-fold coordination) and 0.78~\AA\ (8-fold coordination), 
a change of 14\%. This observation indicates that our analysis does not distinguish between perovskites 
which can be synthesized at ambient or high pressures. Therefore, the prediction of Fe$_2$O$_3$ perovskite
demonstrates that the geometric blueprint can be in general used as a tool for not only predicting
novel perovskite compounds, but also novel high-pressure perovskite phases of previously known
polymorphs.

\newpage
\clearpage

\begin{widetext}
 \begin{figure*}[t!]
 \begin{center}
 \includegraphics[width=0.8\textwidth]{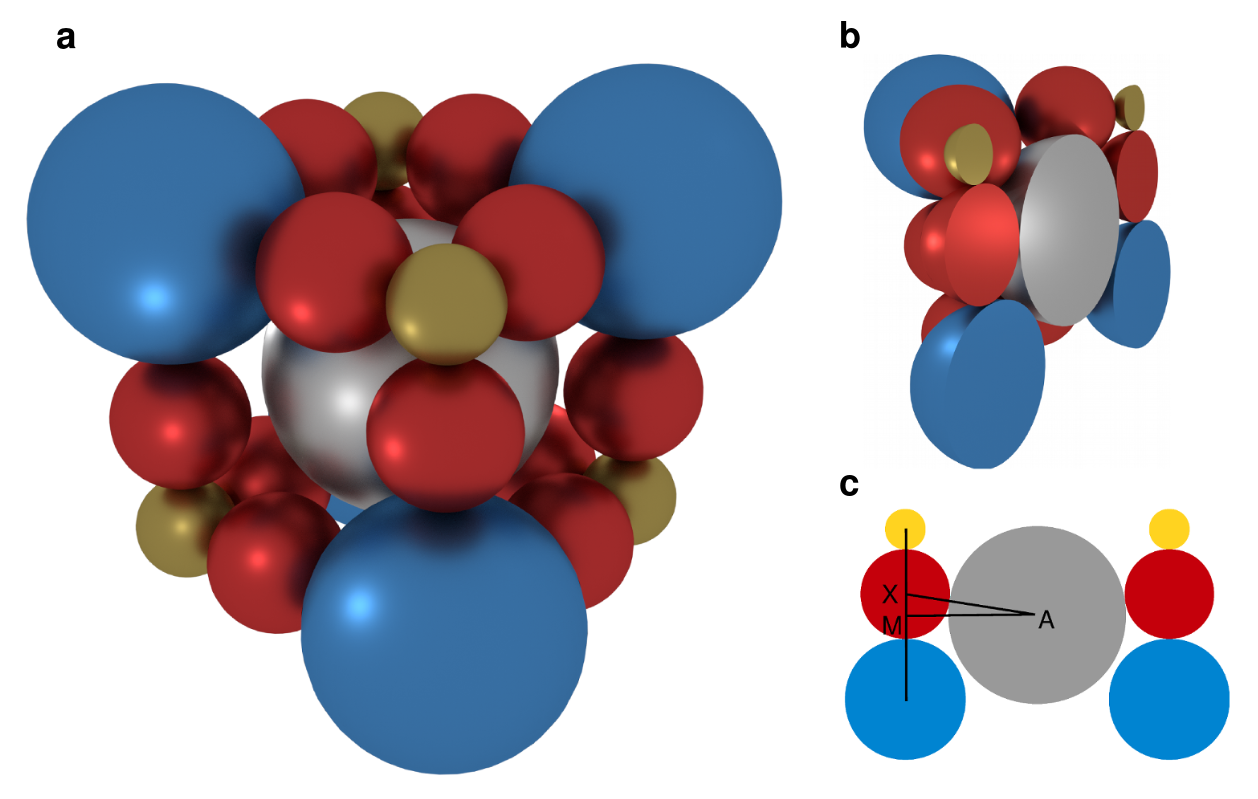}
 \end{center}
 \linespread{1.1}
 \end{figure*}
 {\small 
 \textbf{Figure~S1}. {\bf Rigid sphere model of double perovskites}.
 {\bf a}, Illustration of a cubic double perovskite A$_2$BB$'$X$_6$ in the rigid-sphere model. The 
 A cation is in grey, B and B$'$ are blue and yellow, respectively, and X is in red.
 {\bf b}, Schematic representation of the stretch limit for double perovskites. In this section 
 cutting through the centers of A, B, B$'$, and X we see that the A cation is in contact with the X 
 anion. 
 {\bf c}, The triangle used to derive the stretch limit for double perovskites, in the same 
 configuration as in {\bf b}.
 }
\newpage
\clearpage

 \begin{figure}[t!]
 \begin{center}
 \includegraphics[width=\textwidth]{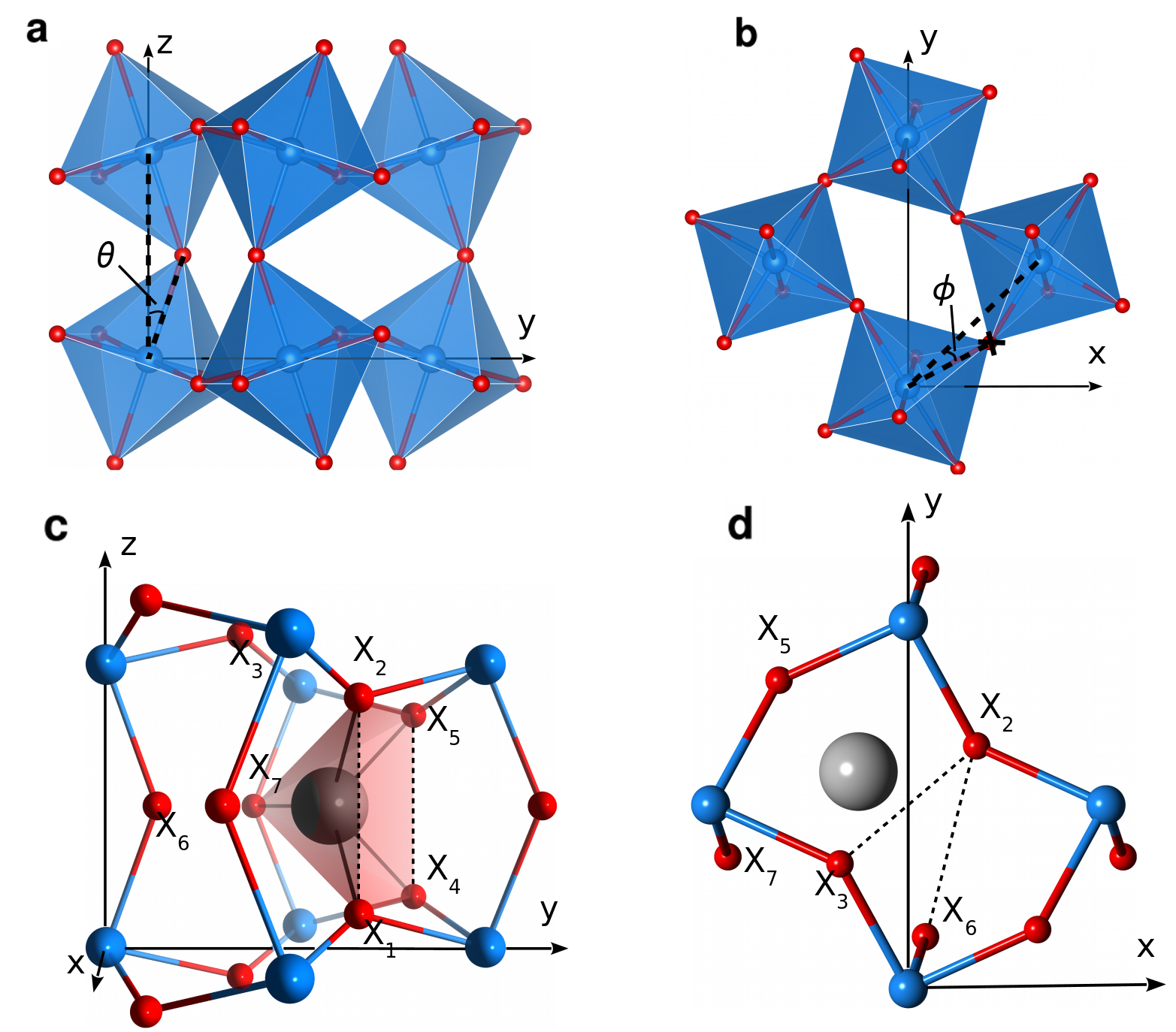}
 \end{center}
 \linespread{1.1} 
 \end{figure}
 {\small
 \textbf{Figure~S2}. {\bf Geometric construction for the tilt limit}. 
 Polyhedral models of the perovskite cavity in the tilt limit, {\bf a}, view along the $x$-axis
 and {\bf b}, along the $z$ axis. The octahedral tilt angle $\theta$ and the precession angle $\phi$ 
 are indicated, and the A-site cation is not represented in {\bf a} and {\bf b} for clarity. 
  The black cross in {\bf b} indicates that the angle is taken between the adjacent B
 cations and the projection of the X anion on the $(x,y)$ plane. 
 {\bf c} and {\bf d}, Ball and stick models of the perovskite cavity in the tilt limit, view along 
 the $x$-axis ({\bf c}) and view along the $z$ axis ({\bf d}). The ions are labelled according to 
 Eqs.~(\ref{eq.1})--(\ref{eq.7}). The dotted lines mark the X anions that come in contact in the 
 tilt limit. The coordination of A is highlighted by the shaded polyhedron in {\bf c}. The Cartesian 
 axes refer to an  orthorhombic unit cell in the $Pnam$ space group. 
 }

\newpage
\clearpage

 \begin{figure*}[t!]
 \begin{center}
 \includegraphics[width=0.5\textwidth]{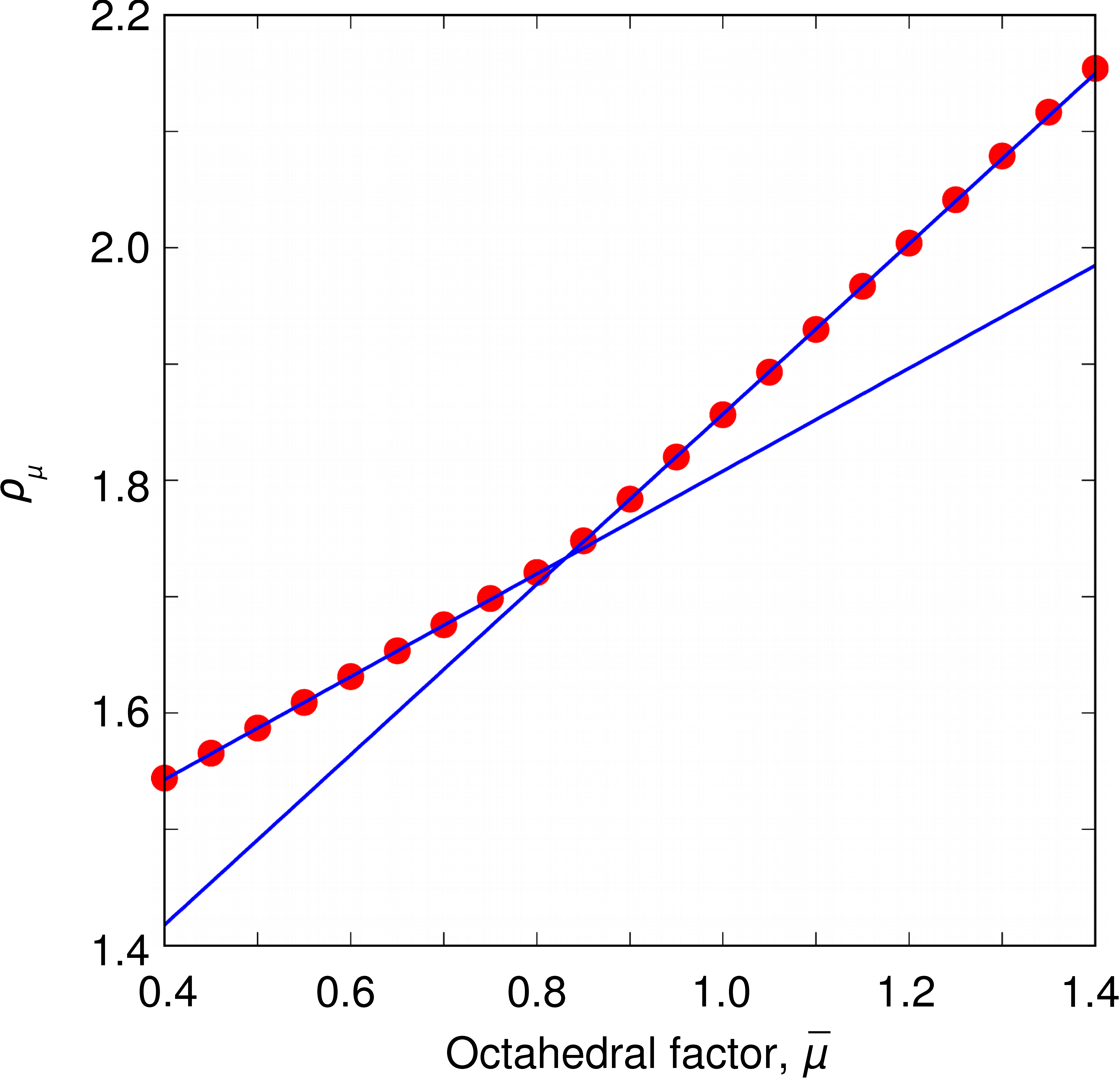}
 \end{center}
 \linespread{1.1}
 \end{figure*}
 {\small
 \textbf{Figure~S3}. {\bf Numerical evaluation of the tilt limit}.
 Plot of the ratio $\rho_\mu=r_{\mu}/r_{\rm X}$ as a function of the average octahedral ratio 
 $\bar{\mu}$, as evaluated  numerically from Eq.~(\ref{eq.16}) for the tilt limits TL1  and TL2 
 (disks). We also show the linear fits of these curves, as reported in Eq.~(\ref{eq11}) (lines).
 }

\newpage
\clearpage

 \begin{figure*}[t!]
 \begin{center}
 \includegraphics[width=\textwidth]{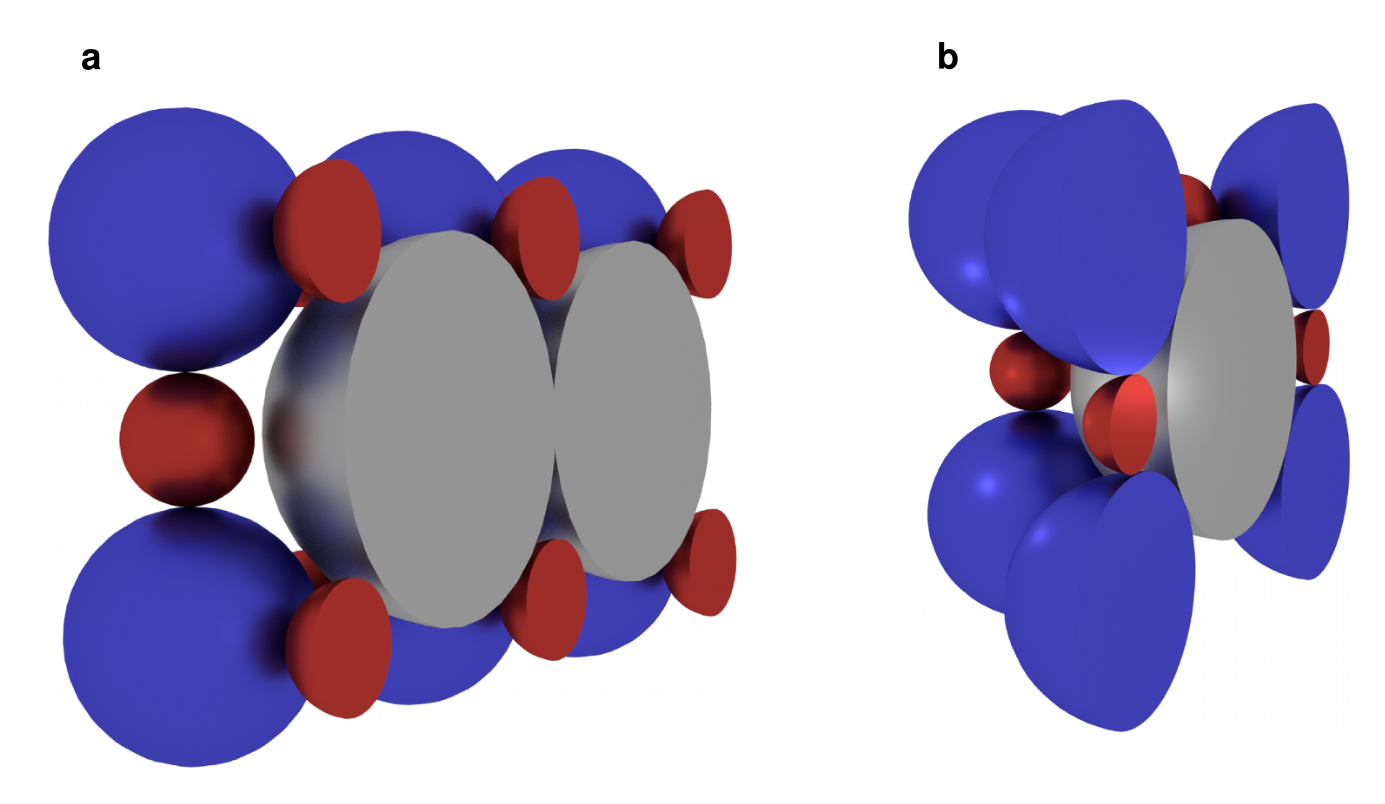}
 \end{center}
 \linespread{1.1}
 \end{figure*}
 {\small
 \textbf{Figure~S4}. {\bf Schematic representation of the secondary stretch limits for
 ABX$_3$ perovskites}.
 {\bf a}, Two adjacent A-site cations are in contact (limit SSL1).
 {\bf b}, The A and B cations are in contact (limit SSL2).  
 The A, B, and X cations are in grey, blue, and red color, respectively.
 }

\newpage
\clearpage

 \begin{figure*}[t!]
 \begin{center}
 \includegraphics[width=\textwidth]{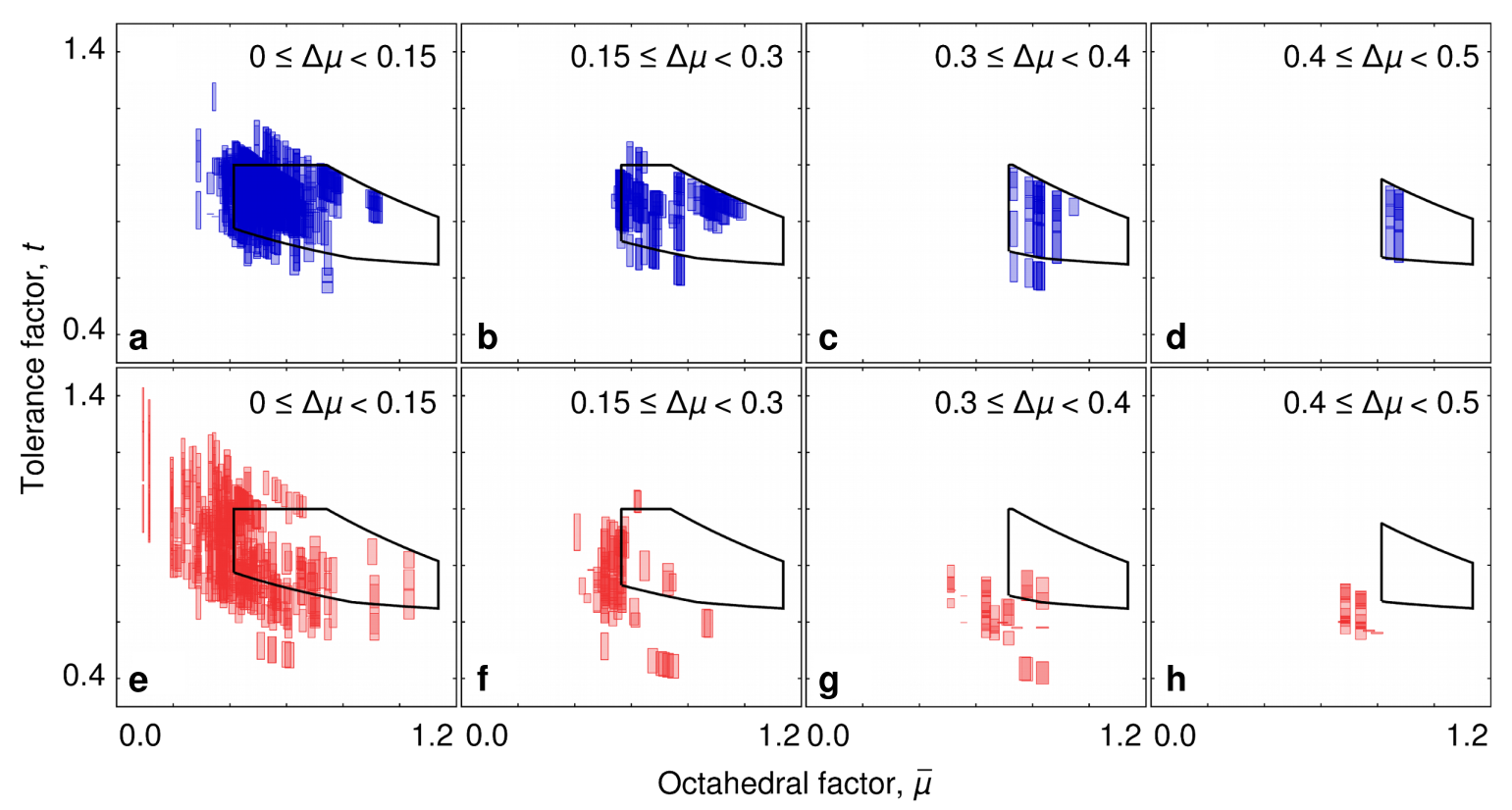}
 \end{center}
 \linespread{1.1}
 \end{figure*}
 {\small
 \textbf{Figure~S5}. {\bf Detailed view of the distribution of known compounds in the
  $\bm{(\bar{\mu}, t, \Delta\mu)}$ space}.
 {\bf a}-{\bf d}, Distribution of known perovskites from databases~S1.1 and S1.3 in the
 $(\bar{\mu}, t)$ plane, for several ranges of the octahedral mismatch $\Delta\mu$
 as indicated at the top of each panel. For each compound we show the two-dimensional
 cross-section of the rectangular cuboid representing the uncertainty in the ionic
 radii. The black lines represent sections of the stability region for $\Delta\mu$ in 
 the middle of the range indicated.
 {\bf e}-{\bf h}, Distribution of compounds from database~S1.2
 which are not perovskites, for several ranges of $\Delta\mu$. The rectangles have the
 same meaning as in panels {\bf a}-{\bf d}. 
 }

\newpage
 \begin{figure*}[t!]
 \begin{center}
 \includegraphics[width=\textwidth]{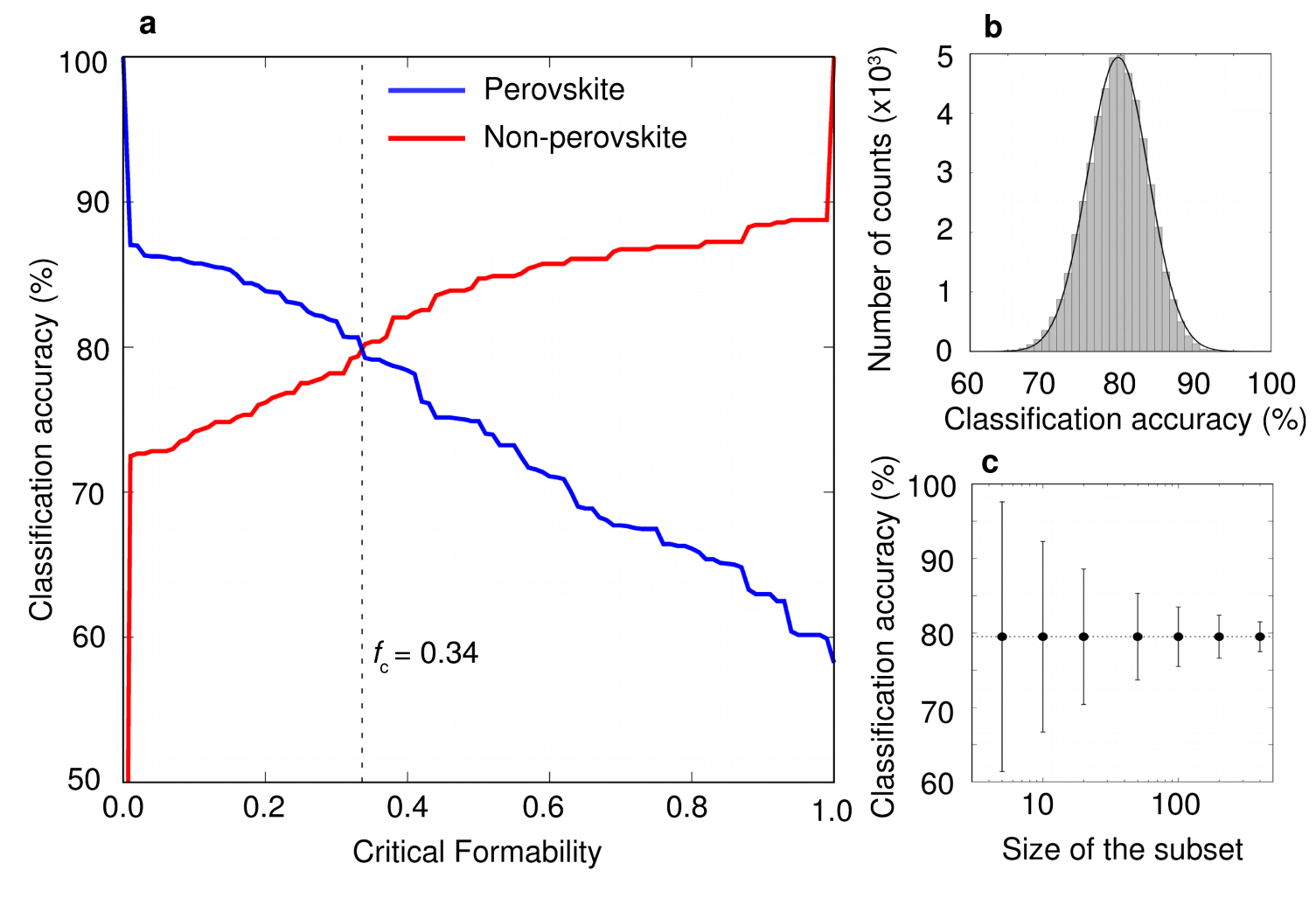}
 \end{center}
 \linespread{1.1}
 \end{figure*}
 {\small
 \textbf{Figure~S6}. {\bf Statistical analysis of the classification accuracy}.
 {\bf a}, Classification accuracy of the geometric model, as a function of the
 critical formability $f_{\rm c}$. The blue line is for perovskites from database~S1,
 the red line is for non-perovskites in the same database. The lines intersect at
 $f_{\rm c}=0.34$, with this choice the classification accuracy is 79.7\%.
 {\bf b}, Analysis of the classification uncertainty. We consider
 50,000 subsets of 100 compounds, randomly drawn from database~S1, and we plot the
 distribution of the classification accuracy of each subset (histogram). We obtain an approximately
 normal distribution, with a standard deviation of 4\%. A Gaussian
 function with the same standard deviation and mean is also shown for comparison.
 {\bf c}, Mean and standard deviation of the classification accuracy calculated as in {\bf b}, 
 for subsets of varying size. The standard deviation decreases for larger subsets.
 }

\clearpage
\newpage
 
 \begin{figure*}[t!]
 \begin{center}
 \includegraphics[width=\textwidth]{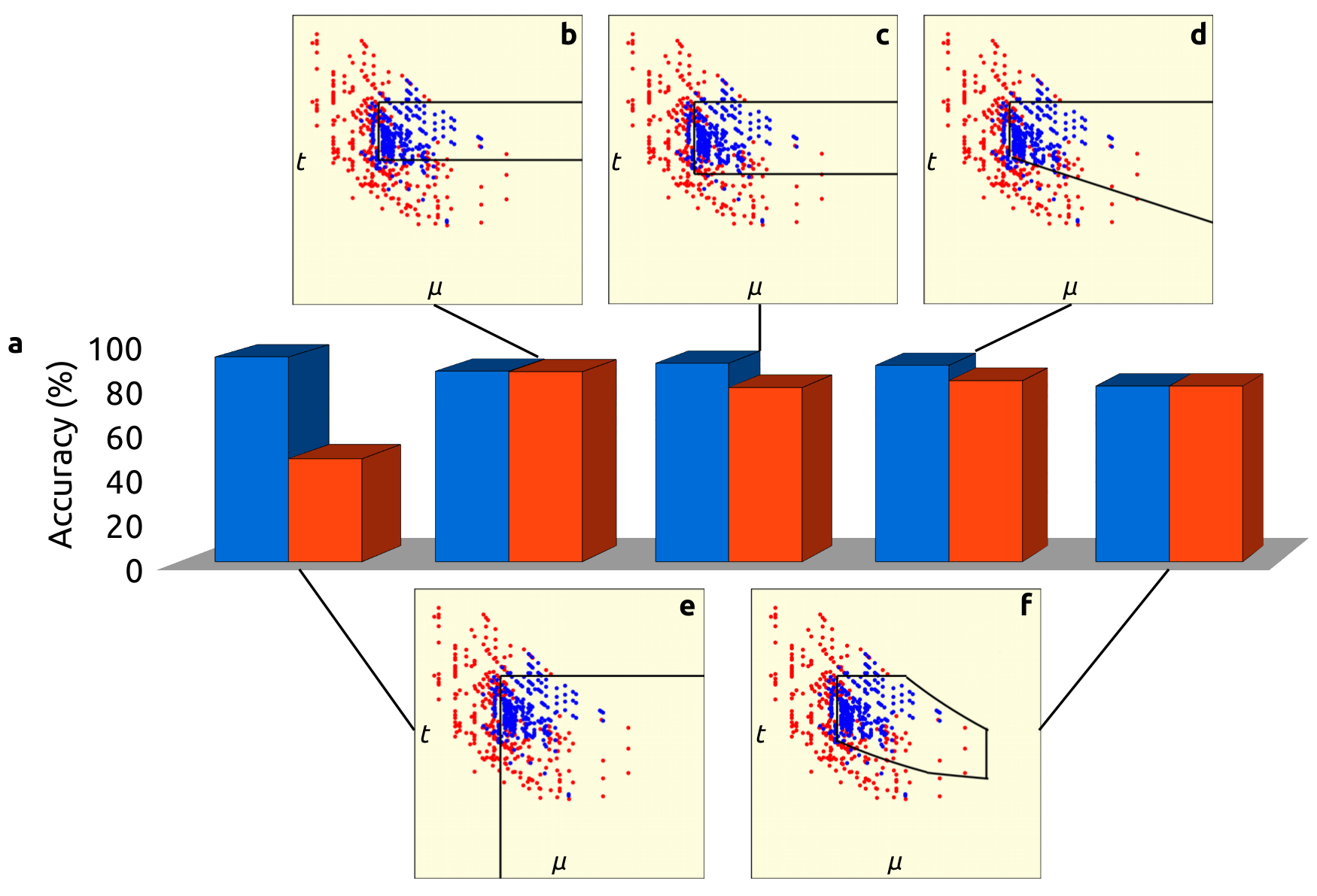}
 \end{center}
 \linespread{1.1}
 \end{figure*}
 {\small
 \textbf{Figure~S7}.  
 {\bf Comparison with other perovskite regions proposed in literature}.
 Accuracy of different perovskite from literature (a): 
 Ref.~\citenum{Giaquinta1994} (b: octahedral limit, stretching limit and $t\geq 0.8$),
 Ref.~\citenum{Navrotsky1998} (c: octahedral limit, stretching limit and $t\geq 0.75$), 
 Ref.~\citenum{Li2004} (d: octahedral limit, stretching limit and $t\geq 0.91-0.23\mu$),
 Goldschmidt's model\cite{Goldschmidt} (e : octahedral limit and stretching limit). (f) 
 corresponds to  the same analysis performed on our model. The bars indicate the accuracy 
 with which a model classifies all compounds in Database~S1.1 as perovskites (blue) and 
 Database S1.2 as non-perovskites (red). For (b-e) we calculate the accuracy using the 
 average $t$, $\bar{\mu}$ and $\Delta \mu$ (as described in the caption of Figure~2 of the 
 main manuscript), while for (f) we take into account the variability of the ionic radii, as
 described in the Supporting Information. The data points in each maps are only shown for 
 ternary compounds for clarity. The red dots correspond to non-perovskites and the blue dots 
 correspond to perovskites.}

\clearpage
\newpage

 \begin{figure*}[t!]
 \begin{center}
 \includegraphics[width=\textwidth]{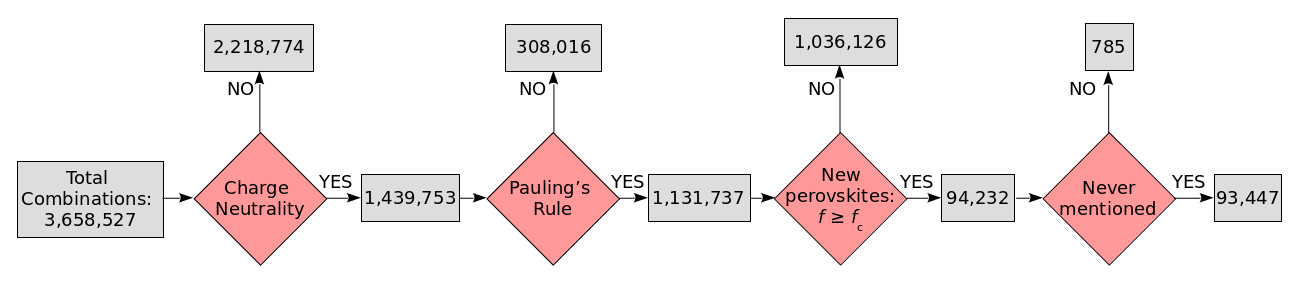}
 \end{center}
 \linespread{1.1}
 \end{figure*}
 {\small
 \textbf{Figure~S8}. {\bf High-throughput screening of perovskites using the geometric model}.
 Schematic flow-chart of the procedure used to predict which compounds, among all 
 3,658,527 hypothetical combinations of A, B, B$'$, and X ions, will crystallize in a
 perovskite or double perovskite lattice. }

\newpage
\clearpage

 \begin{figure*}[t!]
 \begin{center}
 \includegraphics[width=\textwidth]{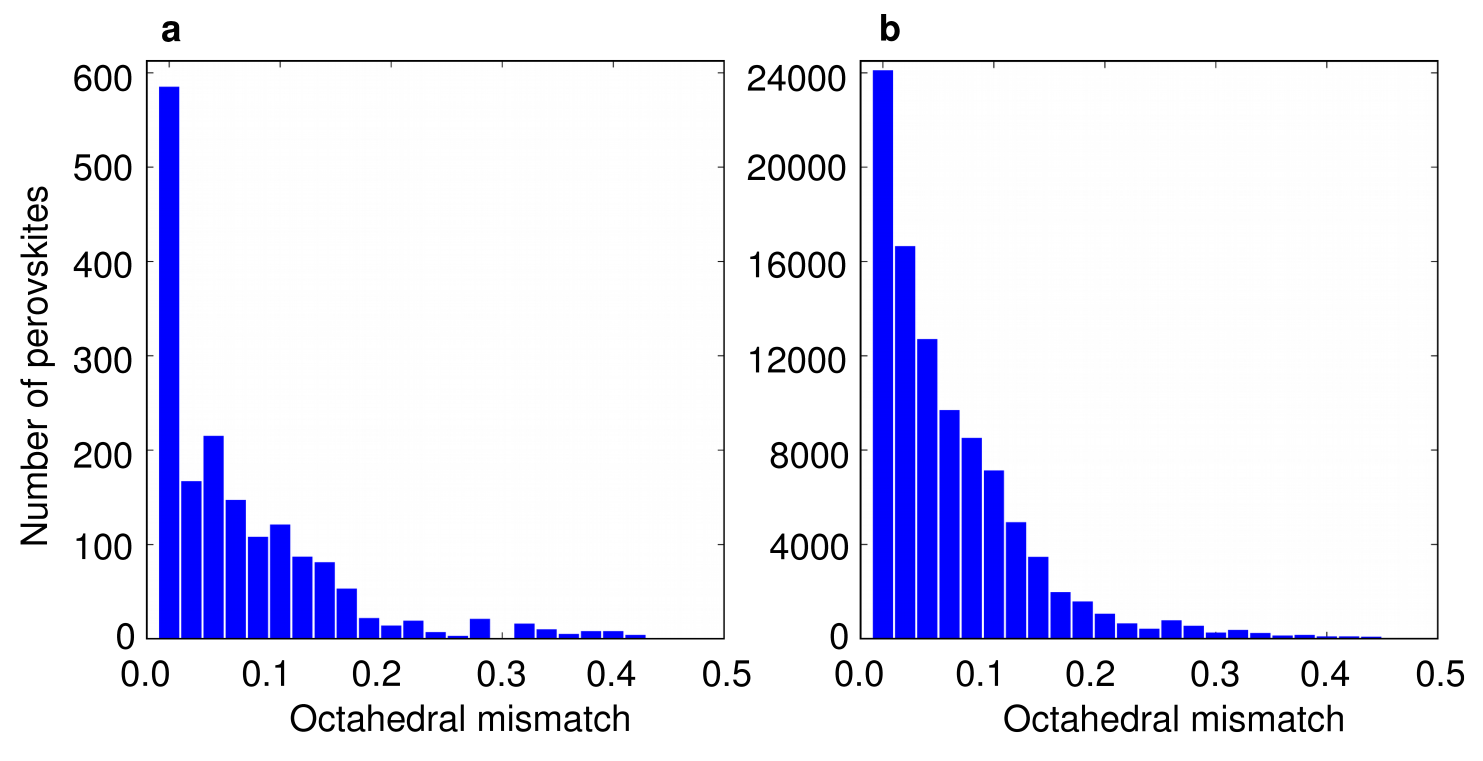}
 \end{center}
 \linespread{1.1}
 \end{figure*}
 {\small
 \textbf{Figure~S9}. {\bf Distribution of known and predicted perovskites as a function of
 octahedral mismatch}.
 {\bf a}, The histogram shows the number of known perovskites from database~S1 (sticks)
 across the range of the octahedal mismatch $\Delta\mu$. The majority of compounds (94\%) 
 is found in the range $\Delta\mu<0.2$. {\bf b}, Same analysis as in {\bf a}, but for the 
 predicted perovskites from database~S2 (sticks). Also in this case the majority of 
 perovskites (96\%) is found in the range $\Delta\mu<0.2$.
 }

 \newpage
 \clearpage

 \begin{figure*}[t!]
 \begin{center}
 \includegraphics[width=\textwidth]{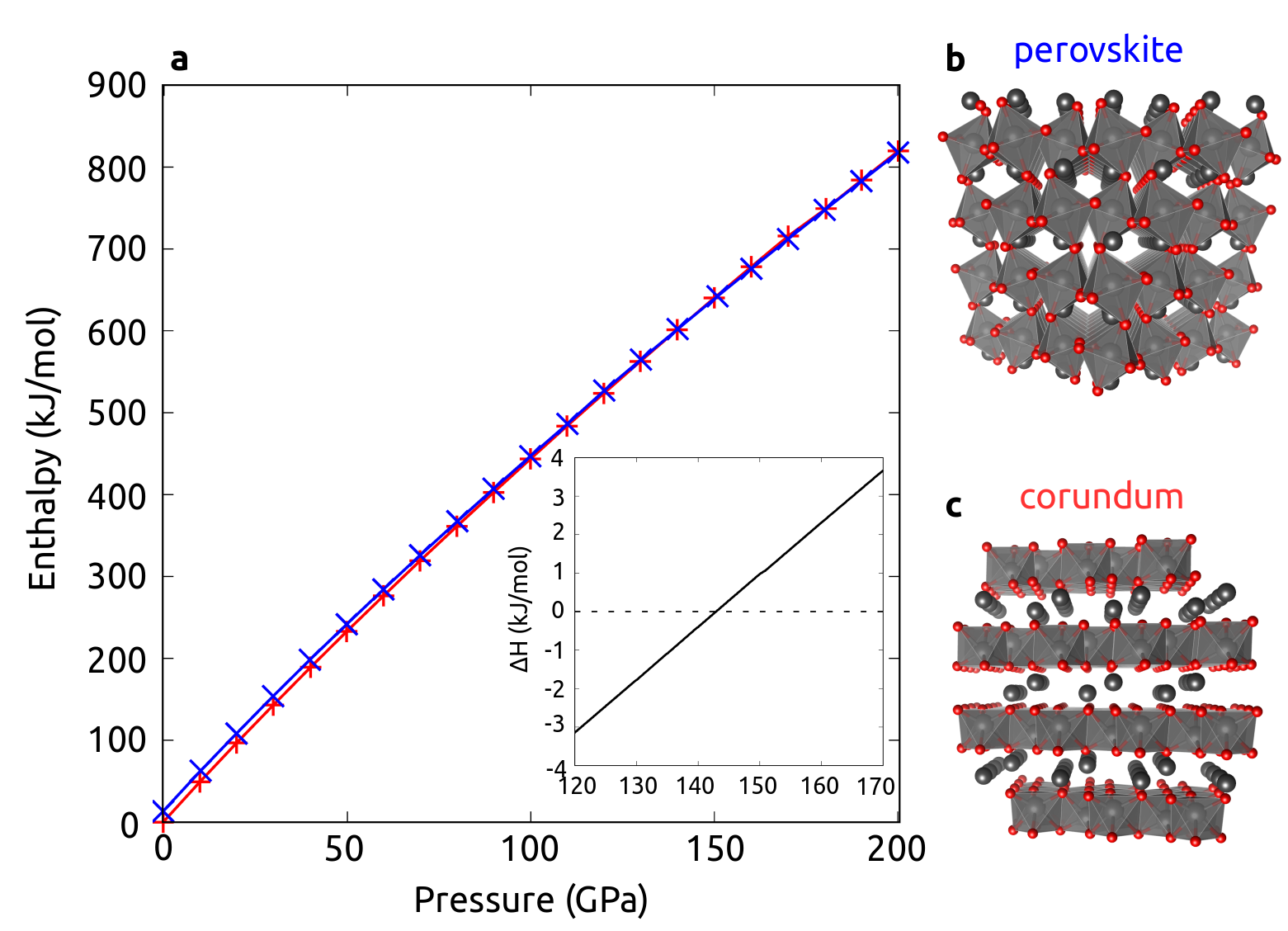}
 \end{center}
 \linespread{1.1}
 \end{figure*}
 {\small
 \textbf{Figure~S10}.
 {\bf Calculated enthalpies for the perovskite (blue) and corundum 
 phases (red) Fe$_2$O$_3$ as a function of pressure}. 
 The enthalpies are referred to the total energy of Fe$_2$O$_3$ corundum at 0 GPa. The inset
 shows the difference in the enthalpies of the corundum and perovskite phases as a function of
 pressure ($\Delta H = H_{\rm hematite} - H_{\rm perovskite}$). Enthalpies are calculated from the
 local density approximation to density functional theory, DFT/LDA, by performing full structural
 optimizations at pressures between 0 and 200 GPa (with 10 GPa increments). Calculations are
 performed using ultrasoft pseudopotentials, a plane wave cutoff of 60~Ry,  a charge density cutoff
 of 300 Ry and a $\mathbf k$-point grid of $4\times4\times4$ centered at $\Gamma$. The lines in the
 (a) are obtained by interpolating through the data points, and the black line in the inset is
 obtained as the difference of the interpolated enthalpies. ({\bf b}-{\bf c}) Polyhedral models of
 the perovskite and corundum crystal structures. The grey spheres correspond to the Fe ions and the
 red spheres correspond to O. The structures were obtained by replacing Ti with Fe in the FeTiO$_3$
 ilmenite and perovskite structures, respectively.
 } 
\end{widetext}

\end{document}